\documentclass[12pt]{article}

\usepackage{cite,psfig}

\usepackage{amssymb}
\usepackage{graphicx}

\def\be{\beta} 
\def\ga{\gamma}
\def\de{\delta}

\def\La{\Lambda}

\def\bM{{\mathbf{M}}}

\def\bW{{\mathbf{W}}}

\def\K{{\mathcal{K}}}

\def\W{{\mathcal{W}}}

\newcommand{\ben}{\begin{equation}}
\newcommand{\een}{\end{equation}}
\newcommand{\bea}{\begin{eqnarray}}
\newcommand{\eea}{\end{eqnarray}}
\newcommand{\ba}{\begin{array}}
\newcommand{\ea}{\end{array}}
\newcommand{\bit}{\begin{itemize}}
\newcommand{\eit}{\end{itemize}}

\def\math{\mathsurround 0pt}
\def\oversim#1#2{\lower.5pt\vbox{\baselineskip0pt \lineskip-.5pt
        \ialign{$\math#1\hfil##\hfil$\crcr#2\crcr{\scriptstyle\sim}\crcr}}}
\def\lap{\mathrel{\mathpalette\oversim {\scriptstyle <}}}

\def\pa{\partial}

\def\half{\frac{1}{2}}

\begin{document}
\thispagestyle{empty}

\def\thefootnote{\fnsymbol{footnote}}

\begin{flushright}
IEM-FT-215/01\\
IFT-UAM/CSIC-01-20\\
SUSX-TH/01-027
\end{flushright}

\vspace{1cm}

\begin{center}

{\Large\sc {\bf BPS Domain Walls in super Yang--Mills \\ and 
Landau--Ginzburg models }}\\[3.5em]
{\large B.~de~Carlos, M.B.~Hindmarsh, N.~McNair}

\vspace*{0.4cm}

{\sl Centre for Theoretical Physics, University of Sussex, \\ 
Falmer, Brighton BN1 9QJ, United Kingdom }

\vspace*{1cm}

{\large J.M.~Moreno}

\vspace*{0.4cm}

{\sl Instituto de Estructura de la Materia, CSIC \\
Serrano 123, 28006 Madrid, Spain }

\end{center}

\vspace*{2.5cm}

\begin{abstract}

We study domain walls in two different extensions of super Yang--Mills 
characterized by the absence of a logarithmic term in their
effective superpotential. The models, defined by the usual gaugino condensate
and an extra field $Y$, give different patterns
of domain walls despite both leading to the same effective
limit for heavy $Y$, i.e. the Veneziano--Yankielowicz effective
Lagrangian of super Yang--Mills. We explain the origin of those differences 
and also give a physical motivation for introducing the field $Y$.

\end{abstract}

\def\thefootnote{\arabic{footnote}}
\setcounter{page}{0}
\setcounter{footnote}{0}

\newpage

\section{Introduction}     
     
The dynamics of strongly coupled theories is a very interesting topic
which has been the subject of detailed investigation for many years.    
It is now believed that the presence of supersymmetry (SUSY) helps
in improving our understanding of these theories, by giving us a
better insight into issues such as non-perturbative effects and the
dynamics of confinement. In that respect SUSY would be not so much a 
phenomenologically relevant symmetry but a tool through which we could
improve our knowledge of the structure of the theory.
     
In particular supersymmetric Yang--Mills theory (SYM), the theory of 
gluons and their SUSY partners, the gluinos, offers a very attractive 
feature, namely the possibility of obtaining exact results despite 
it being a strongly coupled theory.
In 1996 Dvali and Shifman~\cite{Dvali97} studied the formation of
domain walls in this model and estimated the energy of such
objects when the walls were BPS-saturated (in which case also one half
of the supersymmetry is preserved on the wall). Domain walls arise because 
SYM has $T({\rm G})$ degenerate vacua, where $T({\rm G})$ is the
Dynkin index for the group G in the adjoint
representation~\footnote{From now on we will consider 
${\rm G}={\rm SU}(N)$, therefore  $T({\rm G})=N$.}. At some scale 
$\Lambda$ the theory enters a strong coupling regime and gaugino 
condensates form, breaking the $Z_{2N}$ symmetry down to $Z_2$. The
gaugino condensate has been studied in detail for many years and by 
several groups using different methods~\cite{Witte82}. Given 
$\lambda_a$, the gaugino field, we define the condensate as
\begin{equation}
\langle \lambda \lambda \rangle \equiv \langle {\rm Tr} \lambda^a 
\lambda_a \rangle = \Lambda^3 e^{i \frac{2\pi}{N} k} \;\;\;, 
k=0,\ldots,N-1 \;\;.
\label{cond}
\end{equation}
However if one wants to study SYM any further, it is necessary
to find an effective description in terms of composite superfields.
This was done in the early Eighties by Veneziano and 
Yankielowicz (VY)~\cite{Venez82} and the resulting Lagrangian is
\begin{equation}
{\cal L} = \frac{1}{4} \int d^4 \theta {\cal K} + \frac{1}{2}
\left[ \int d^2 \theta {\cal W} + {\rm h.c.} \right] \;\;,
\label{lag_susy}
\end{equation}
where
\begin{equation}
{\cal K} = (S \bar{S})^{1/3} \;\;,
\label{Kahler}
\end{equation}
is the K\"ahler potential and      
\begin{equation}
{\cal W} = \frac{2}{3} S \left( \ln \left( \frac{S}{\Lambda^3} 
\right)^N -N \right) \;\;,
\label{supVY}
\end{equation}
is the superpotential, where $S$ is the composite chiral superfield
whose lowest component is the gaugino condensate
\begin{equation}
S \equiv \frac{3}{32 \pi^2} \langle {\rm Tr} (\omega_{\alpha} 
\omega^{\alpha}) \rangle  = \frac{3}{32 \pi^2}  \langle {\rm Tr} 
(\lambda_{\alpha} \lambda^{\alpha}) \rangle + \ldots
\label{comp}
\end{equation}
The structure of ${\cal W}$ is uniquely determined by the anomaly and
the symmetries of the theory. Although this Lagrangian is perfectly
adequate to reproduce the vacuum structure of SYM, two major
problems arise when one tries to use it to study the dynamics of the
gaugino condensates. The first one is the lack of $Z_{N}$ invariance 
of the superpotential (\ref{supVY}), which was pointed out by Kovner 
and Shifman~\cite{Kovne97} a few years back\footnote{Note that, in
terms of the $S$ superfield, the invariance of the theory is $Z_N$ 
rather than the original $Z_{2N}$.}. The second problem is related to 
the presence of the logarithm in Eq.~(\ref{supVY}), which results in
the multivaluedness of the scalar potential. This problem was 
addressed (and solved) in the context of the linear multiplet 
formulation of gaugino condensation with a field dependent 
coupling~\cite{Burge95}. In the chiral formulation, the
concept of the {\em glued potential} was proposed in Ref.~\cite{Kovne97},     
and further developed in Ref.~\cite{Kogan98}, to cure both problems.
However, it was pointed out in this latter reference how the presence
of {\em cusps} (points where the effective potential is continuous
but non-differentiable) between the different vacua of the model would
constitute a problem when trying to construct domain walls.

It therefore seems that the VY approach is somehow incomplete and we
should be trying to incorporate new degrees of freedom in the effective
description of SYM. Several proposals along this line have been
put forward in previous years, and a certain degree of success has
been achieved in applying them to the construction of domain walls. Promoting
the scale parameter $\Lambda$ to a dynamical field (the glueball order 
parameter) was one of them~\cite{Gabad99}, which resulted in the 
construction of BPS-saturated domain walls in the large-$N$
limit~\cite{Dvali99} by exploiting the similarity of these models to a
certain kind of Landau--Ginzburg construction. Another option is to
incorporate chiral matter fields to the VY Lagrangian, something that
was already considered in the Eighties by Taylor, Veneziano and
Yankielowicz (TVY)~\cite{Taylo83}. The idea is to study this model in
its Higgs phase (i.e. for small values of the masses of the matter
fields) and try to recover the VY limit when this mass parameter gets 
large. A lot of work has been devoted to following this approach in
the context of building domain walls~\cite{Smilg97,Decar99,Decar01,Binos01,
Smilg01}. In particular~\cite{Decar99} it is possible to construct 
BPS-saturated domain walls for any value of the mass parameter (and, 
therefore, obtain an effective VY limit) only when the number of 
matter flavours, 
$N_f$, is less than $N/2$ and, moreover, the logarithmic branch of 
the superpotential is crossed if $N_f > 1$. Although ways were found 
of circumventing these obstructions~\cite{Decar01,Smilg01}, the
presence of the logarithm with its different possible branches is still not
totally understood, and it would be desirable to find an alternative 
formulation where no logarithms were present other than in the limit
were the VY theory was recovered. This is precisely the goal of this
paper. A more ambitious objective would be to identify the extended 
models and new fields we introduce with already known 
physical systems. In particular, there has been a lot of work in
recent years devoted to the connection between BPS-saturated domain
walls and branes, starting with the work of Witten~\cite{Witte97},
and we shall explore whether any connections are possible between
our constructions, done in the low-energy field theory, and other
studies performed in the context of string or M-theory~\cite{Vafa00}, 
and also in the presence of higher supersymmetries~\cite{Kaplu99,Dorey00}.

In Section~\ref{s:models} we present two interpolating models, which 
we denote as Landau--Ginzburg-like, with their different interesting 
limits, one of which is precisely VY. In Section~3 we start by 
presenting a simple example, with $N=3$, in order to illustrate the 
different aspects of the calculation, followed by a detailed
discussion of more general results obtained in both models. In 
Section~4 we motivate them from the physical point of view, and in 
Section~5 we study the large-$N$ expansion in the VY effective limit.
We conclude in Section~6. We have also included a pair of
Appendices, one explaining the numerical methods used to construct 
the domain walls, and the other supplying a detailed analysis of
the BPS constraint equations.

\section{Interpolating models}
\label{s:models}
In this section we present two model supersymmetric field theories 
which interpolate between the Veneziano--Yankielowicz (VY) effective 
Lagrangian for supersymmetric Yang--Mills theory (SYM), and the 
so-called $A_N$ Landau--Ginzburg models.

Let us consider the following superpotential which, from now on, will
be referred to as Model~I,
\begin{equation}
{\cal W} (S, Y) =  S \left( \left(\frac{\Lambda^{3}}{S}\right)^N 
e^{-Y/M+N-1} + \frac{Y}{M} \right) \;\;,
\label{e:W1}
\end{equation}
with $\Lambda$ and $M$ two mass parameters. We will assume a canonical
K\"ahler potential for the dimension one fields $\phi = S^{1/3}$ and Y,
i.e.
\begin{equation}
{\cal K} (S, \bar{S}, Y, \bar{Y}) = (S \bar{S})^{1/3} + Y \bar{Y} \;\;.
\label{e:kahler}
\end{equation}
Notice that the superpotential transforms as ${\cal W} \rightarrow 
e^{i \, 2\pi/N} \, {\cal W}$ when $ S\rightarrow   e^{i \, 2\pi/N} \,
S$, which generates a $Z_{N}$ symmetry in the Lagrangian. The 
supersymmetric vacua of this theory are given by solutions to the 
equations $\W_S=0$ and $\W_Y=0$, where
\bea
\W_S &=& (1-N)\left(\frac{\Lambda^{3}}{S}\right)^N e^{-Y/M+N-1} + 
  \frac{Y}{M} \;\;, \label{e:W_S}\\
\W_Y &=& \frac{S}{M} \left( -\left(\frac{\Lambda^{3}}{S} \right)^N 
e^{-Y/M+N-1} + 1 \right) \;\;.
\label{e:W_Y}
\eea
There are $N$ such solutions,
\ben
Y^a_* = M(N-1), \quad
S^{a}_* =  \Lambda^{3}e^{i 2\pi a/N},
\een
with $a = 0,1,\dots,N-1$, which spontaneously break the $Z_N$ symmetry. 
In these minima, the superpotential is given by $ \W_{a} = N S^{a}_* $. 

The mass eigenstates are most easily found in terms of the $\phi$ and 
$Y$ fields. The mass matrix $\bM$ is given by $\bM = \bW \bW^\dagger$, 
where $\bW$ is the matrix of second derivatives of the superpotential
\ben
\bW = 
\left(\begin{array}{cc} 
\W_{\phi\phi} & \W_{\phi Y}\\
\W_{Y\phi} & \W_{YY}
\end{array}\right).
\een
In the vacuum $(S_{*}^{0},Y_*)$, 
\begin{equation}
\bW_{*}^{0} = 
\left(\begin{array}{cc} 
9N(N-1)\Lambda & 3N\Lambda^2/M \\
3N\Lambda^2/M      & \Lambda^3/M^2
\end{array}\right).
\end{equation}
In the limits of small and large  $\Lambda/M$, one can
straightforwardly show that
\begin{eqnarray}
~m^2_Y \to {\Lambda^6}/{M^4(N - 1)^2},&\quad& m^2_\phi
\to 81N^2(N-1)^2 \Lambda^2 \quad(\La\to 0) \nonumber\\
&  & \label{e:m2largeL} \\
m^2_Y \to \Lambda^6/M^4, 
\phantom{N - 1)^2}
&\quad& m^2_\phi \to 81N^2\Lambda^2
\phantom{N - 1)^2}
\quad(\La\to \infty) \;\;. \nonumber
\end{eqnarray}
Hence
\ben
\frac {m_Y}{ m_{\phi}}  \sim {\cal O}  \left[ \frac {\Lambda}{M} 
\right]^2 \;\;.
\een
{}From now on we will work in $ M = 1 $ units. Then, for large 
values of $\Lambda$, $Y$ is a heavy field. We can therefore integrate 
it out by imposing $\W_Y = 0$, where $\W_Y$ is given in Eq.~(\ref{e:W_Y}).
Substituting the resulting solution for $Y$ in terms of $S$ into the 
superpotential (\ref{e:W1}) we get the VY superpotential of 
Eq.~(\ref{supVY}), up to an unimportant constant factor, 
\begin{equation}
{\cal W}_{\rm VY}
(S) =  S  \left( \ln \frac {S^N} {\Lambda^{3 N}} - N \right) \;\;.
\label{e:WVY}
\end{equation}
On the other hand, if we assume  $\Lambda \ll 1$, then $S$ is a heavy 
field, and we could impose $\W_S=0$ to obtain
\begin{equation}
{\cal W}_{\rm eff} (Y) =  N \Lambda^3\left(\frac{Y}{N-1}\right)^{1-1/N}
 e^{-Y/N+1-1/N} \;\;.
\end{equation}
The resulting potential has a single minimum at $Y=N-1$, as is
expected, which means that the trajectories described by $Y$ start and 
end up at the same point. Moreover, it can be checked that, in the
limit $\La \to 0$, the field barely moves from that point. Therefore
it is a good approximation to freeze $Y$ at its vacuum value, to
obtain an effective superpotential
\ben
\W_{\rm LG1} = S\left( \left(\frac{\La^3}{S}\right)^N + N - 1
\right) \;\;.
\label{e:LG1}
\een
The effective theory now has a form similar to a $Z_N$ symmetric
Landau--Ginzburg model (hence the notation), although it is not usual 
to see negative powers of the field. 

Alternatively, one can consider the following superpotential (our,
from now on, Model~II),
\begin{equation}
{\cal W} (S, Y) =  S \left( 
\left( \frac{S}{\Lambda^3} \right)^N e^{Y/M-N-1} - \frac{Y}{M} \right) \;\;,
\label{e:W2}
\end{equation}
which can be obtained from Eq.~(\ref{e:W1}) by just changing 
($ N \rightarrow - N, \, Y \rightarrow  -Y $ ). Hence, there are also 
$N$ minima
\ben
Y^a_* = N+1, \quad
S^{a}_* =  \Lambda^{3}e^{i 2\pi a/N} \;\;,
\een
with $a = 0,1,\dots,N-1$, which spontaneously break the $Z_N$ symmetry. 
The superpotential in these minima is given by $ \W_* = - N S_* $.
 It is easy to check that the same hierarchy between
the $S$ and $Y$ masses appears in the large/small $\Lambda$ limit. 
On top of that, there is also a $Z_N$ conserving minimum given by
$( Y_* = 0,  S_* = 0)$. Whereas for large $\Lambda$ we get the same
VY limit of Eq.~(\ref{e:WVY}), at small values of $\Lambda$ the model
approaches a conventional Landau--Ginzburg model
\ben
\W_{\rm LG2} = S\left( \left(\frac{S}{\La^3}\right)^N - N - 1\right).
\label{e:LG2}
\een  
Hence we have come up with two different two-field models which, in the 
limit $\La \gg 1$ coincide with the VY Lagrangian, without the 
ambiguity of a multivalued logarithm and, as far as solutions to the 
field equations for $S$ go, become Landau--Ginzburg models in the limit 
$\La \ll 1$.

\section{Construction of the domain walls}
\label{s:dwconstruct}

Let us study the domain wall spectrum of these models. As we are considering 
static solutions, we wish to minimize the static surface energy functional
\ben
\sigma = \frac{1}{2} \int dz \left( 
\K_{S\bar{S}}\partial_z S\partial_z\bar{S} + 
\K_{Y\bar{Y}}\partial_z Y\partial_z\bar{Y} + V(S,Y,\bar{S},\bar{Y})
\right) \;\;,
\een
where $\K_{J\bar{J}}=\partial^2 \K/ \partial J \partial \bar{J}$,
$J=S,Y$ is the corresponding K\"ahler metric. From now on 
we assume that the wall spreads along the $xy$ plane and, 
therefore, the profile is calculated along the $z$ direction. The
scalar potential $V$ is given by
\ben
V = \K^{-1}_{S\bar{S}}|\W_S|^2 
+ \K^{-1}_{Y\bar{Y}}|\W_Y|^2 \;\;.
\een
As is well known
~\cite{Abraham, Cvetic91}, this can be rewritten as 
\ben
\sigma = \frac{1}{2} \int dz \left(
\K_{S\bar{S}}\left|\pa_z \bar{S} - \K^{-1}_{S\bar{S}}e^{i\ga}{\W}_{S} 
     \right|^2 +
\K_{Y\bar{Y}}\left| \pa_z \bar{Y} - \K^{-1}_{Y\bar{Y}}e^{i\ga}{\W}_{Y}
\right|^2 \right) + \left. {\rm Re} \; 
[e^{i\ga}\W]\right|_{-\infty}^{+\infty} \;\;,
\een
where $\ga$ is an arbitrary constant. Consider a domain wall interpolating 
between vacuum $a$ and vacuum $b$. If we assume that it is BPS, it
will be described by the following equations
\begin{eqnarray}
{\cal K}_{S\bar{S}} \partial_z \bar{S} & = & e^{i\gamma}
\frac{\partial {\cal W}}{\partial S} \;\;, \nonumber \\
     & & \label{TVY_bps} \\
{\cal K}_{Y\bar{Y}} \partial_z \bar{Y} & = & e^{i\gamma}
\frac{\partial {\cal W}}{\partial Y} \;\;, \nonumber
\end{eqnarray}
where now $e^{-i\gamma_{ab}}=({\cal W}_b-{\cal W}_a)/|{\cal W}_b-{\cal
W}_a|$, and ${\cal W}_{a(b)}$ is the value of the superpotential
at vacuum $a(b)$. One can easily show that, in that case, the energy
functional saturates the BPS bound
\begin{equation}
\sigma = | {\cal W}_b - {\cal W}_a | \;\;,
\end{equation}     
and     
\ben
\partial_z(e^{i\gamma_{ab}}\W) = V \;\;.
\een
Hence the solution to the BPS equations (if any) interpolates
between  ${\cal W}_a$ and  ${\cal W}_b$  following a 
straight line in the ${\cal W}$-space, obeying the constraint
\begin{equation}
{\rm Im}(e^{i\gamma} {\cal W}(S,Y)) = {\rm constant} \;\;. 
\label{const}
\end{equation}
This constraint equation is the key to classifying the solutions to
the BPS equations.

{}From now on we shall adopt the following parametrization for the
fields $S$ and $Y$
\begin{eqnarray}
     S & = & |S_*| R(z) e^{i \beta(z)} \;\;, \nonumber \\
     &  &  \label{param} \\
     Y & = & {\rm Re} \; [Y](z) + i \, {\rm Im} \; [Y](z) \;\;, \nonumber 
\end{eqnarray}
where $R$ and $\beta$ are real functions. It is therefore easy to see 
that, when going from a vacuum $a$ situated at $z=-\infty$, to another 
one, $b$, which is $k$ neighbours away (i.e. $b=a+k$), situated at
$z=+\infty$, the following boundary conditions apply
\begin{itemize}     
 \item for the $S$ field, 

$R(-\infty)=R(+\infty) = 1$, $\beta(-\infty)= 0$,
$\beta(+\infty) = 2\pi k/N $ 
     
\item for the $Y$ field,

${\rm Re} \; [Y] (-\infty) = {\rm Re} \; [Y] (+\infty) = N \mp 1$, 
${\rm Im} \; [Y] (-\infty) = {\rm Im} \; [Y] (+\infty) = 0$
           
\end{itemize}
while the phase $\gamma$ is given by $\gamma= \mp \pi/2 - \pi k/N$
depending on which model (I or II) we are considering.
We will take $k$ in the range $0 < k < N$. Notice that there
are two possible domain walls connecting two given vacua
corresponding to the two paths defined by $k$ and $N-k$.

\subsection{An example: N=3} 

Before explaining how the different effective limits emerge, it is 
interesting to work out an explicit example. Let us consider the 
simplest case, the extension of SU(3) super Yang--Mills. Since there 
are three vacua, we can always connect any two of them through a 
domain wall that interpolates between nearest neighbours. We want 
to find these domain walls and study their BPS character. In order 
to do that, we have to minimize the energy of a static 
configuration with the required boundary conditions. Our
problem involves two mass scales and, therefore, the discretization 
has to be done in an optimal way to include them both. The details of 
the minimization algorithm and our convergence criteria are given in 
Appendix A. On the other hand, if we assume that the domain wall is 
BPS-saturated, we can try to solve the corresponding first order 
differential equations given by Eq.~(\ref{TVY_bps}). To do that, we 
have to find the appropriate (if any) initial conditions for the fields 
at the centre ($z=0$) of the domain wall. Using a symmetric ansatz 
allows us to fix the phases of the fields at this point. We are then 
left with two parameters, the absolute values of the fields ($R$ and 
${\rm Re} \; [Y]$) but, since they have to obey the constraint 
Eq.~(\ref{const}), only one of them is free. If there is a BPS domain 
wall, we can use the usual overshooting-undershooting method to fix 
that parameter, and then obtain the profile.

We have used both methods. Of course, minimizing the energy gives 
more information since it allows us to explore also non-BPS domain walls. 
We have found that there is a BPS domain wall for all values of 
$\Lambda$, irrespective of which extension we are using. The domain 
wall path in $S$ and $Y$ space for the first model, defined by 
Eq.~(\ref{e:LG1}), can be seen in Figs~1a and 1b respectively, where 
we plot the Argand diagrams for different values of $\Lambda$. Notice 
that they get stabilized both in the small and in the large $\Lambda$ 
limits. 

As we will see, the same applies to the profiles when we plot them as 
a function of the relevant ($z$ or $\Lambda z$ ) variable.
     
\begin{figure}
\centerline{
\includegraphics[width=7.5cm]{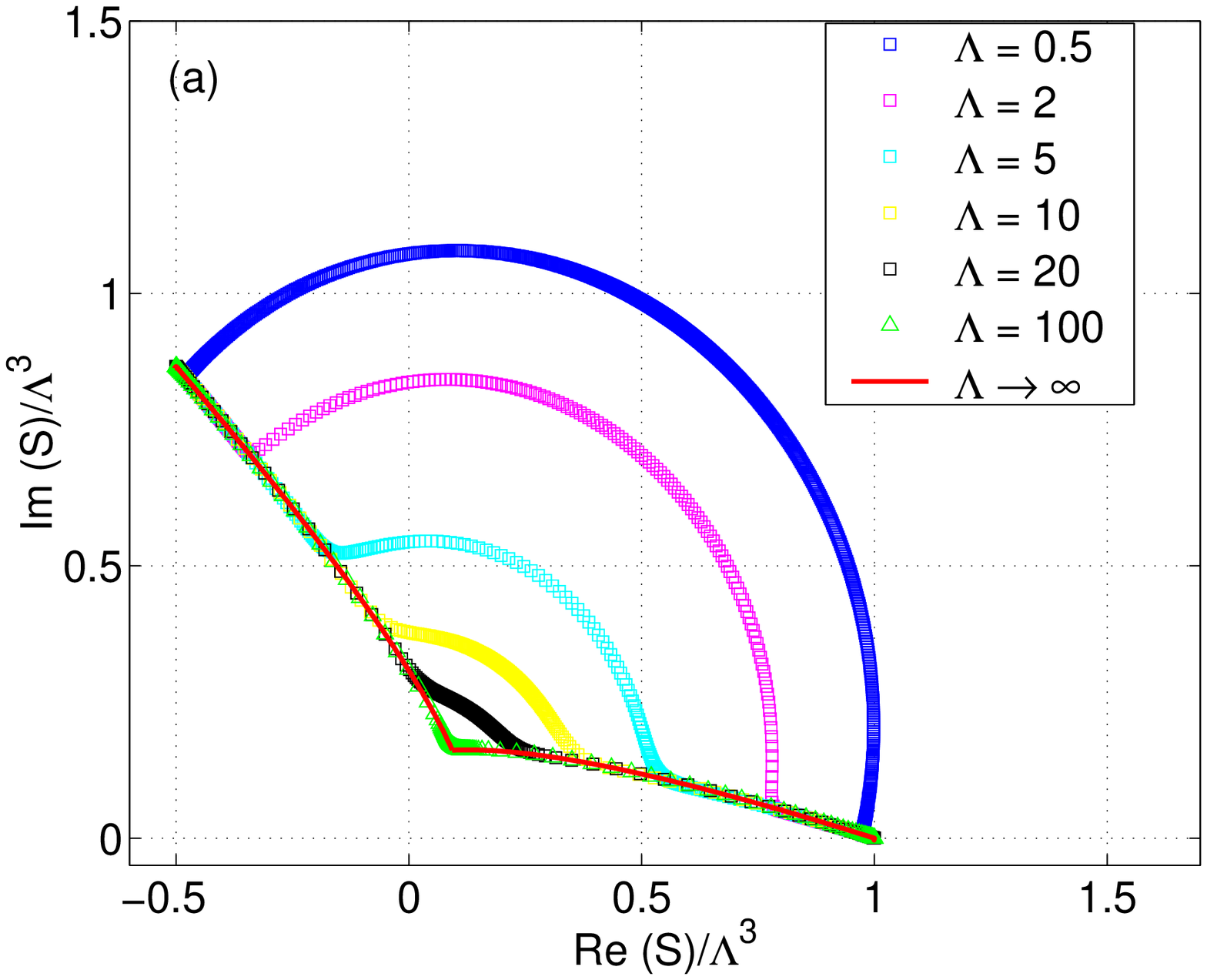}
\includegraphics[width=7.5cm]{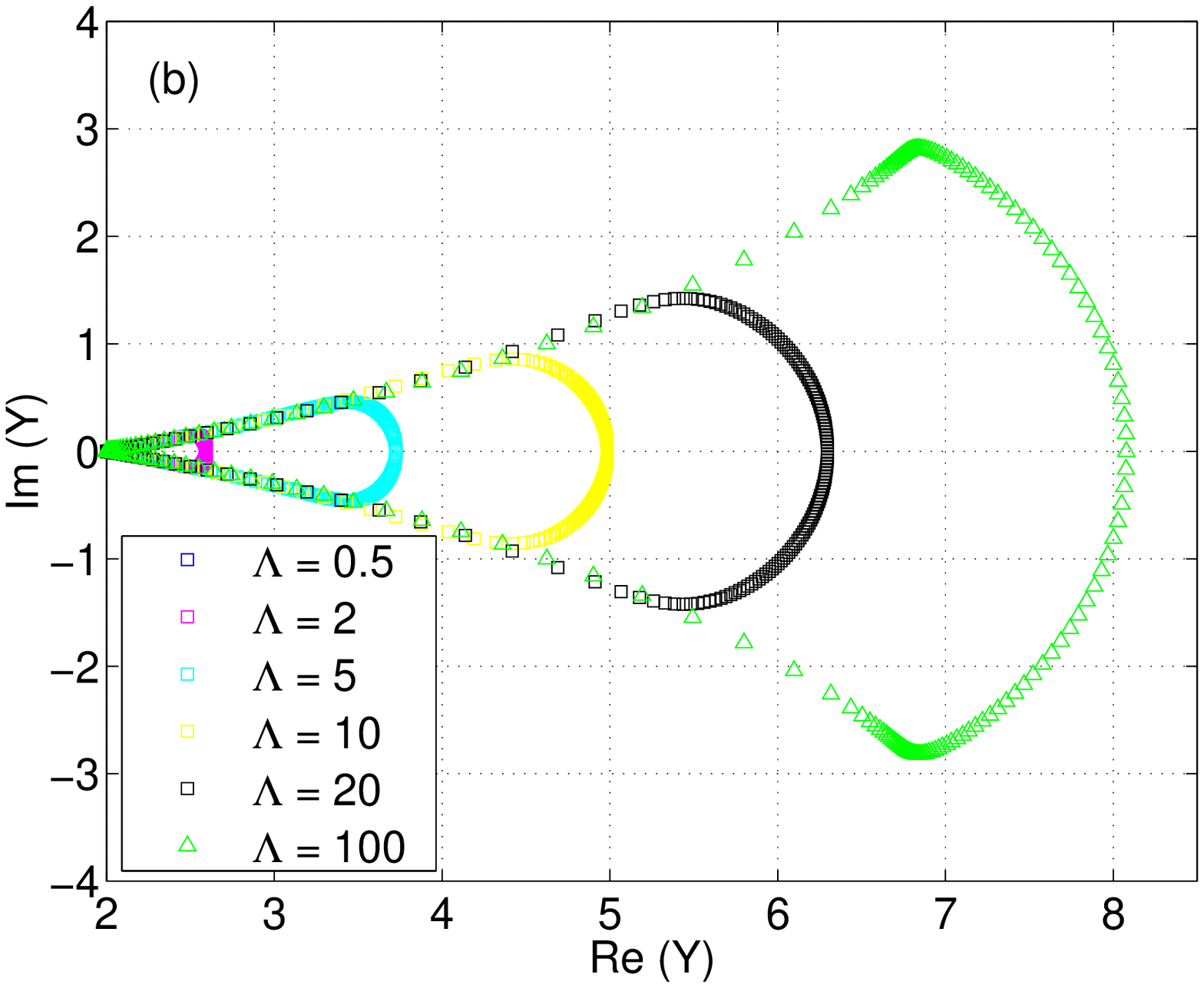} }
\caption{}
{\footnotesize \noindent \textbf
{\bf (a)} Argand diagram of the profiles of the gaugino 
condensate $S$ for $N=3$ and several values of $\Lambda$ in Model~I.
The continuous line represents the effective VY limit; 
{\bf (b)} Argand diagram of the profiles of the $Y$ field for the
same model and values of the parameters as before. }
\label{fig1}
\end{figure}

Let us first focus on the large $\Lambda$ case. The profile for $S$ is 
made out of two branches, $z>0$ and $z<0$. According to our theoretical 
analysis of Section~2, this configuration should be also described by 
an effective VY Lagrangian. In fact, we have checked that it can be 
obtained by solving the BPS equations for the VY Lagrangian with an 
appropriate prescription for the logarithm. In particular, the $z<0$ 
($z>0$) branch is obtained using $ \ln S^N = \ln (\Lambda^3 R)^N + 
N i \arg (S)  \; ( - 2 \pi i )$. The profile for $Y$ can be divided in 
three regions. The central one corresponds to the core, where $S$ 
remains almost constant but $Y$ changes very quickly, as can be seen 
in Figs~2. Notice that the width of this core goes to zero when we 
increase $\Lambda$. Moreover, the $Y$ profile can be calculated from 
the $S$ one. This is because, as expected in this limit, the condition 
$\W_Y =0$ is satisfied almost everywhere (it is only violated in the 
region where the phase is forced to change almost in an abrupt way to 
connect the two $S$ branches, as can be seen in Fig.~2b).

\begin{figure}
\centerline{
\includegraphics[width=7.5cm]{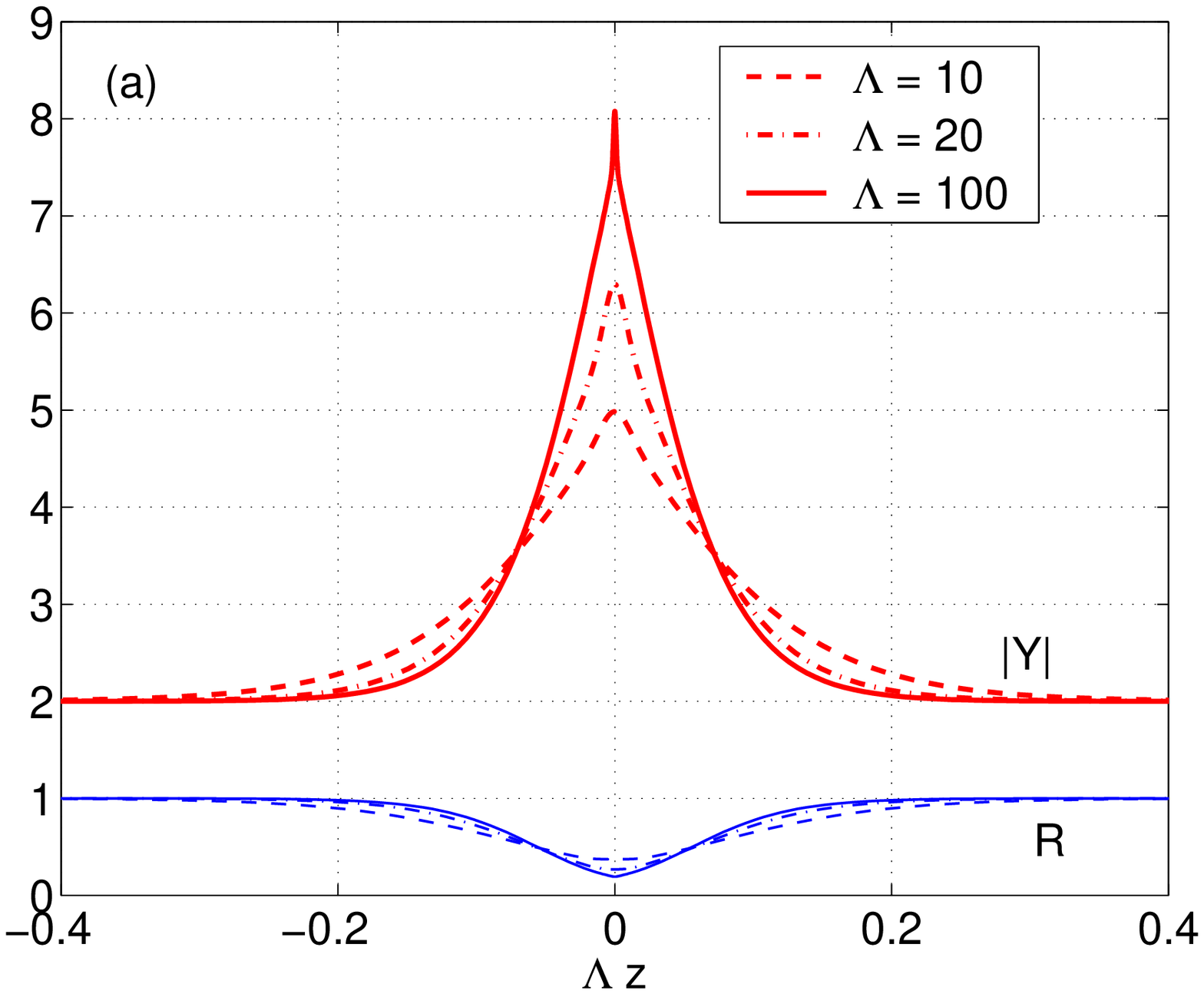}
\includegraphics[width=7.5cm]{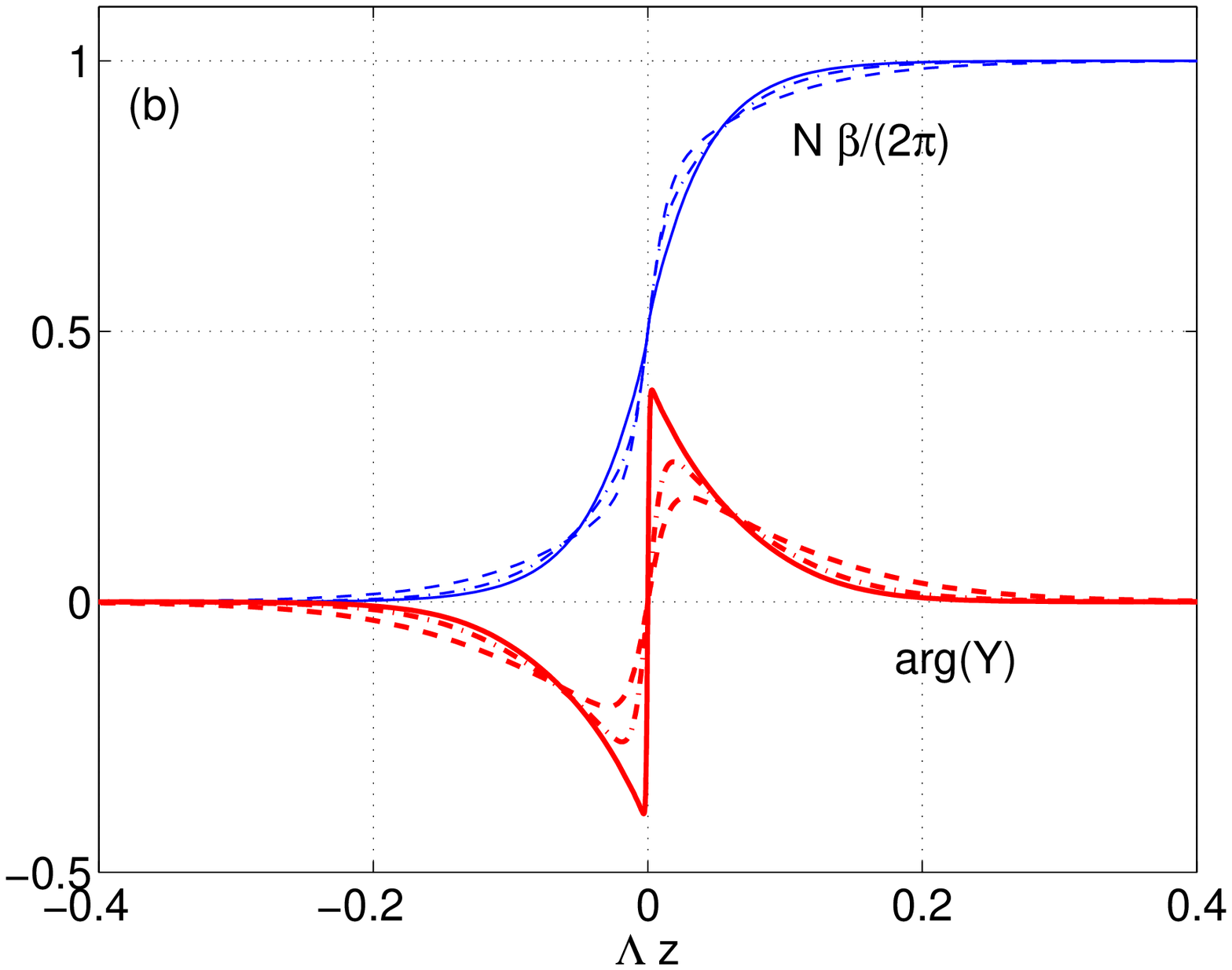} }
\caption{}
{\footnotesize \noindent \textbf
{\bf (a)} Profiles of $R$ (absolute value of $S$) and $|Y|$ 
as a function of $\Lambda z$ for $N=3$ and large values of
$\Lambda$ in Model~I (solid: $\Lambda=100$, dot-dashed: 
$\Lambda=20$, dashed: $\Lambda=10$); {\bf (b)} profiles of 
$N \beta/(2 \pi)$ and ${\rm arg}(Y)$ for the same model and 
values of the parameters as before. }
\label{fig2}
\end{figure}

In the small $\Lambda$ limit, the $Y$ field is frozen to its value at 
the minimum, $N \mp 1$ depending on the model we are considering (as
can be clearly seen in Fig.~1b). We have checked that the profile for 
the $S$ field coincides with the corresponding BPS domain walls in the 
effective Landau--Ginzburg models of Eqs~(\ref{e:LG1},\ref{e:LG2}). This 
is illustrated in Figs~3, where we show both the absolute value $R$ 
(Fig.~3a) and the phase $\beta$ (Fig.~3b) of the $S$ field in Model~I
as a function of the rescaled position, $\Lambda z$, for small values of 
$\Lambda$, as well as its theoretical $\Lambda \rightarrow 0$ limit. 
     
\begin{figure}
\centerline{
\includegraphics[width=7.5cm]{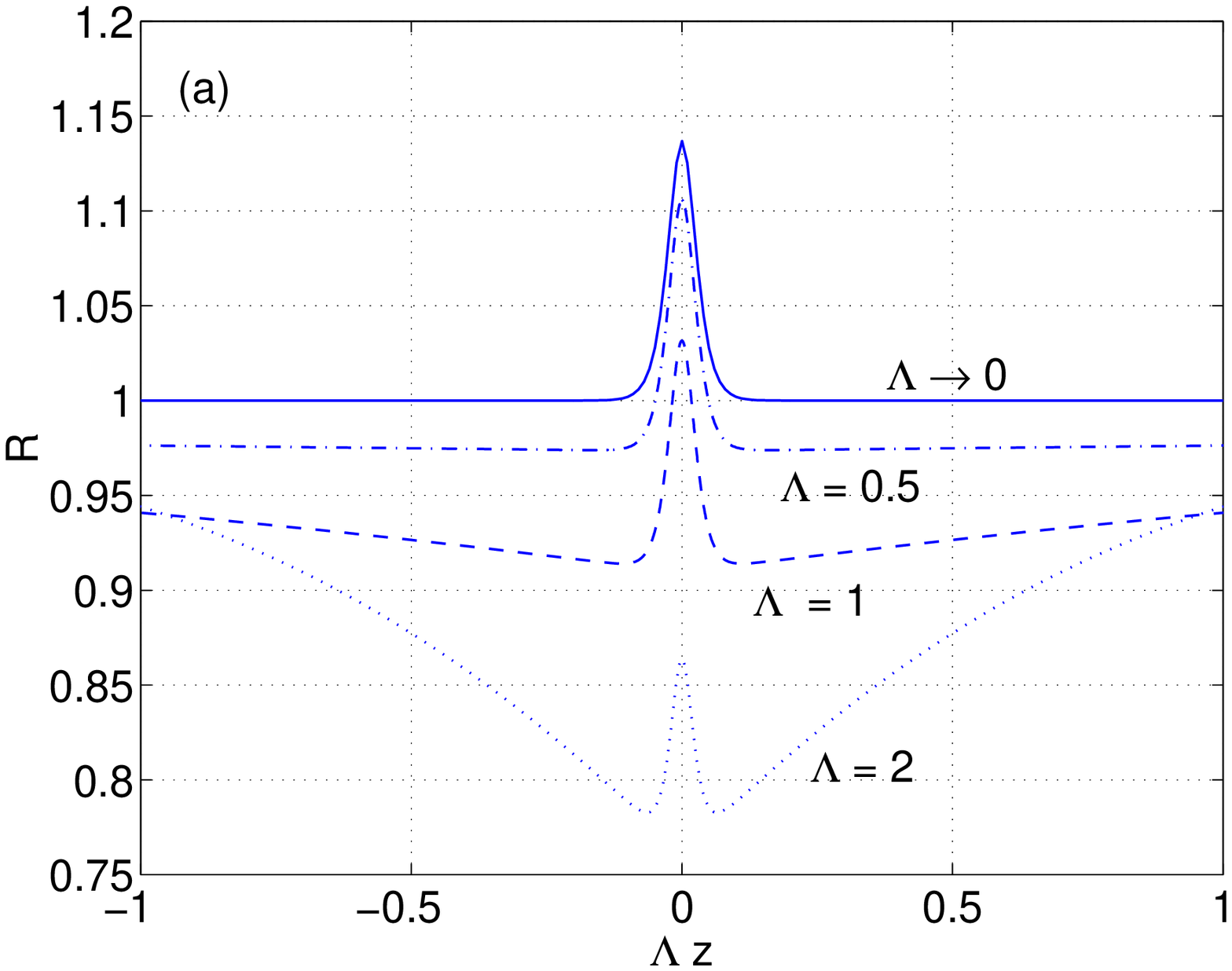}
\includegraphics[width=7.5cm]{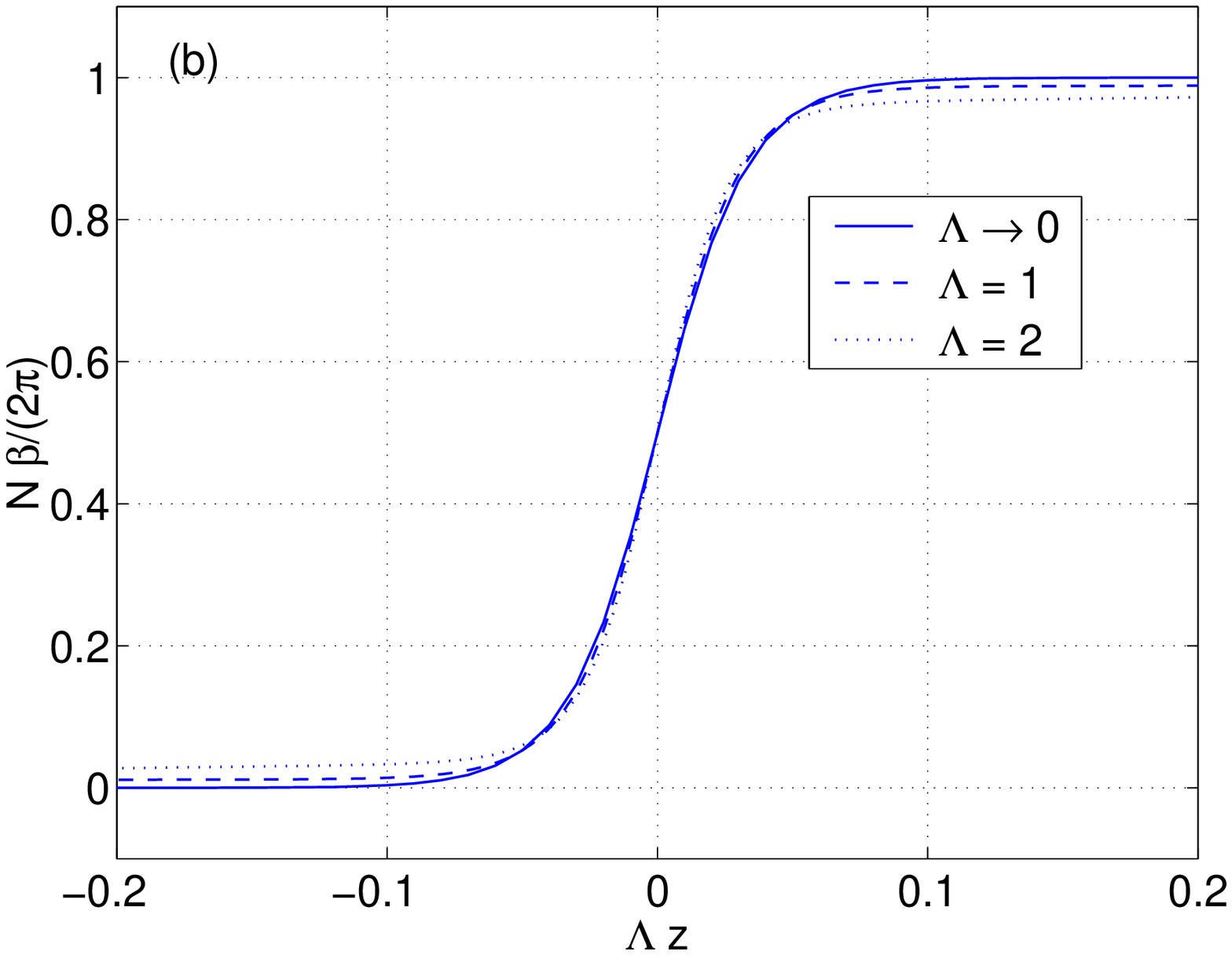} }
\caption{}
{\footnotesize \noindent \textbf
{\bf (a)} Profiles of $R$ (absolute value of $S$), as a function of 
$\Lambda z$, for $N=3$ and small values of $\Lambda$ (dotted: 
$\Lambda=2$, dashed: $\Lambda=1$, dot-dashed: $\Lambda=0.5$, solid: 
LG limit $\Lambda \rightarrow 0$) in Model~I; {\bf (b)} 
profiles of $N \beta/(2 \pi$) for the same model and 
values of the parameters as before. The $\Lambda=0.5$ curve is not shown
explicitly as it is indistinguishable from the LG limit on the graph.}
\label{fig3}
\end{figure}

We have therefore found a uniparametric family of BPS domain walls 
for the case $(N=3, k=1)$. The theory describing these walls 
interpolates between the super Yang--Mills Veneziano--Yankielowicz 
model (with some given rules about how to glue the different sectors 
of the potential) and a Landau--Ginzburg model. 
On the other side, we have also analyzed the $(N=3, k=2)$ case.
There are no BPS domain walls with these boundary conditions 
(see Appendix \ref{a:c-anal}), in other words the domain wall 
connects the two vacua through the shortest path in $S$ space.

Let us now try to extend this result for higher values of $N$, where a 
richer pattern of domain walls is possible. Notice that BPS domain 
walls have to satisfy the constraint given by Eq.~(\ref{const}). This 
equation involves two fields which, in the limits explained before 
($\Lambda \rightarrow \infty$ and $\Lambda \rightarrow 0$), should 
split into two simple -more restrictive- equations, since the dynamics 
are controlled only by one field. As we will see, the compatibility of 
these equations is the key to understanding the pattern of 
domain walls for arbitrary $N$ values.

\subsection{General analysis: results}
     
When considering domain walls in models with $N>3$, the possibility 
of having different types of walls, connecting the different vacua,
arises. We first consider complex walls, which connect different 
vacua with $|S| = \La^3$, and briefly discuss real walls (connecting 
the vacuum in Model II at $|S|=0$ to a vacuum at $|S| = \La^3$) at 
the end of the section. 

For the moment we shall exclude the cases where $k=N/2$ (with $N$ 
even). They correspond to  $S(-\infty) = - S(\infty)$ and give a 
constraint that includes $S=0$ as a possible point in their 
trajectory, where we have seen that the two models we are 
considering (I and II) have radically different behaviour. For 
a detailed study of these cases the reader is referred to
Appendix~\ref{a:c-anal}.
     
In our numerical searches for both BPS and non-BPS solutions, we have 
found just one domain wall solution interpolating between two given 
vacua. Our conclusions from these results and from our analysis of the 
BPS constraint equations detailed below are that, for finite $\Lambda$,
\begin{itemize}     
     
\item the domain walls found in Model~I are BPS-saturated {\em only} 
for $k=1$;
     
\item in Model~II we have found BPS-saturated walls for {\em any}
value of $k$ in the range $1 \leq k < N/2$. This actually covers 
all possible situations given that, for any $N$, there is always a 
path with $k<N/2$ that links any two chirally asymmetric vacua. 

\end{itemize}
These statements are proved in Appendix~\ref{a:c-anal} by going to 
the limit $\Lambda \to 0$ and appealing to the continuity of the 
solutions as a function of $\Lambda$. Moreover they are illustrated 
in Figs~4 for $N=5$ and $k=2$, where we plot, in Fig.~4a, the Argand 
diagram for the superpotential 
Eq.~(\ref{e:W1}) of Model~I. As we can see the larger $\Lambda$ is, 
the more the curve tends to a straight line, which corresponds to a 
BPS-saturated solution. Therefore, it looks likely that, in the 
$\Lambda \rightarrow \infty$ (VY) limit, these domain walls of Model~I 
will turn into BPS ones.  
     
In Figs.~4b,c we turn to Model~II and plot the profiles of the fields 
$R$ (Fig.~4b) and ${\rm Re} \; [Y]$ (Fig.~4c) as a function of
$\Lambda z$, for several values of $\Lambda$. The behaviour of the 
different curves is very similar to that of the $N=3$, $k=1$ case 
plotted before, and we have also appended in Fig.~4b the 
effective $\Lambda \rightarrow 0$ (Landau--Ginzburg) and $\Lambda 
\rightarrow \infty$ (VY) limits, which we shall discuss in a moment. 

\begin{figure}
\centerline{
\includegraphics[width=7.5cm]{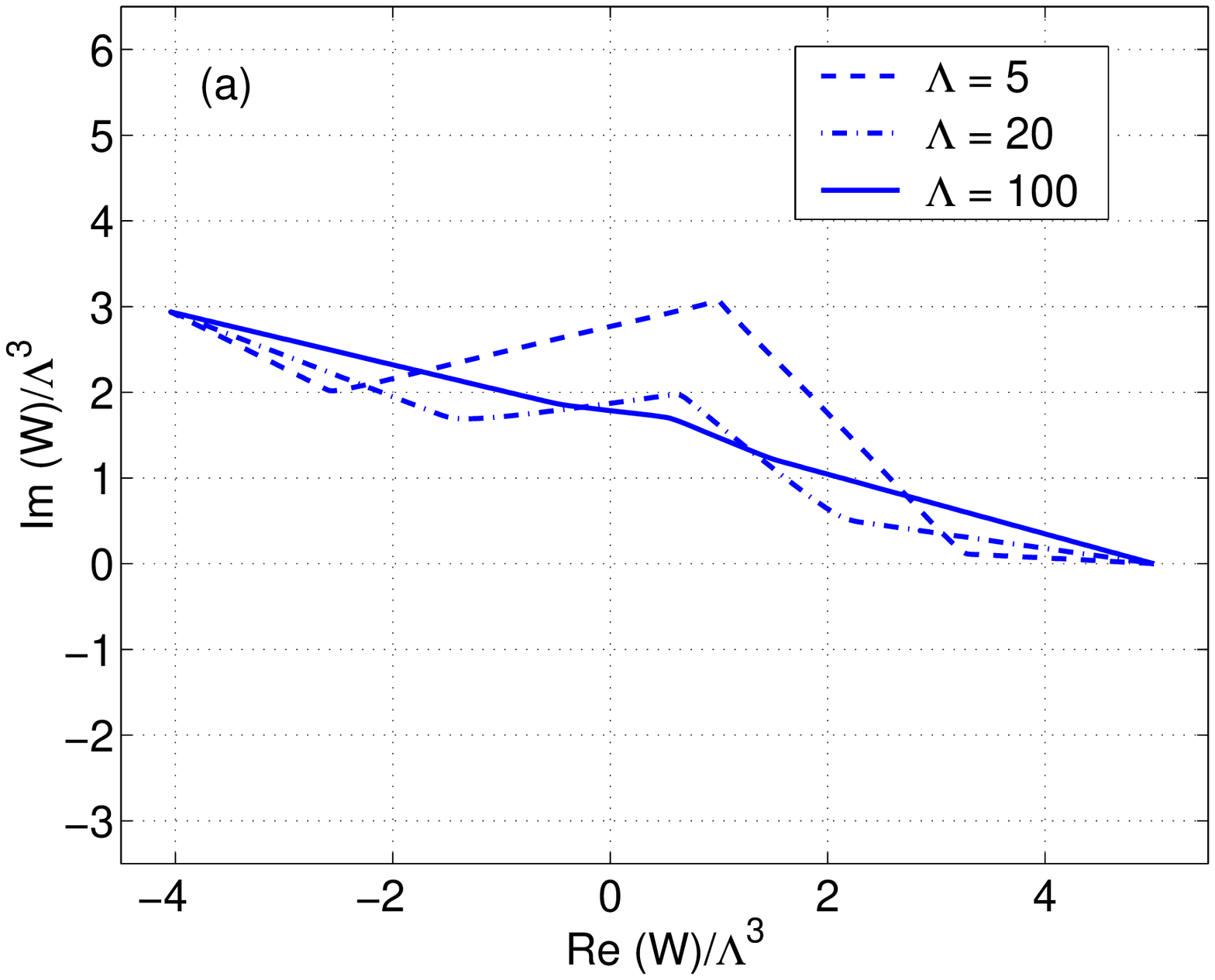}}    
\centerline{
\includegraphics[width=7.5cm]{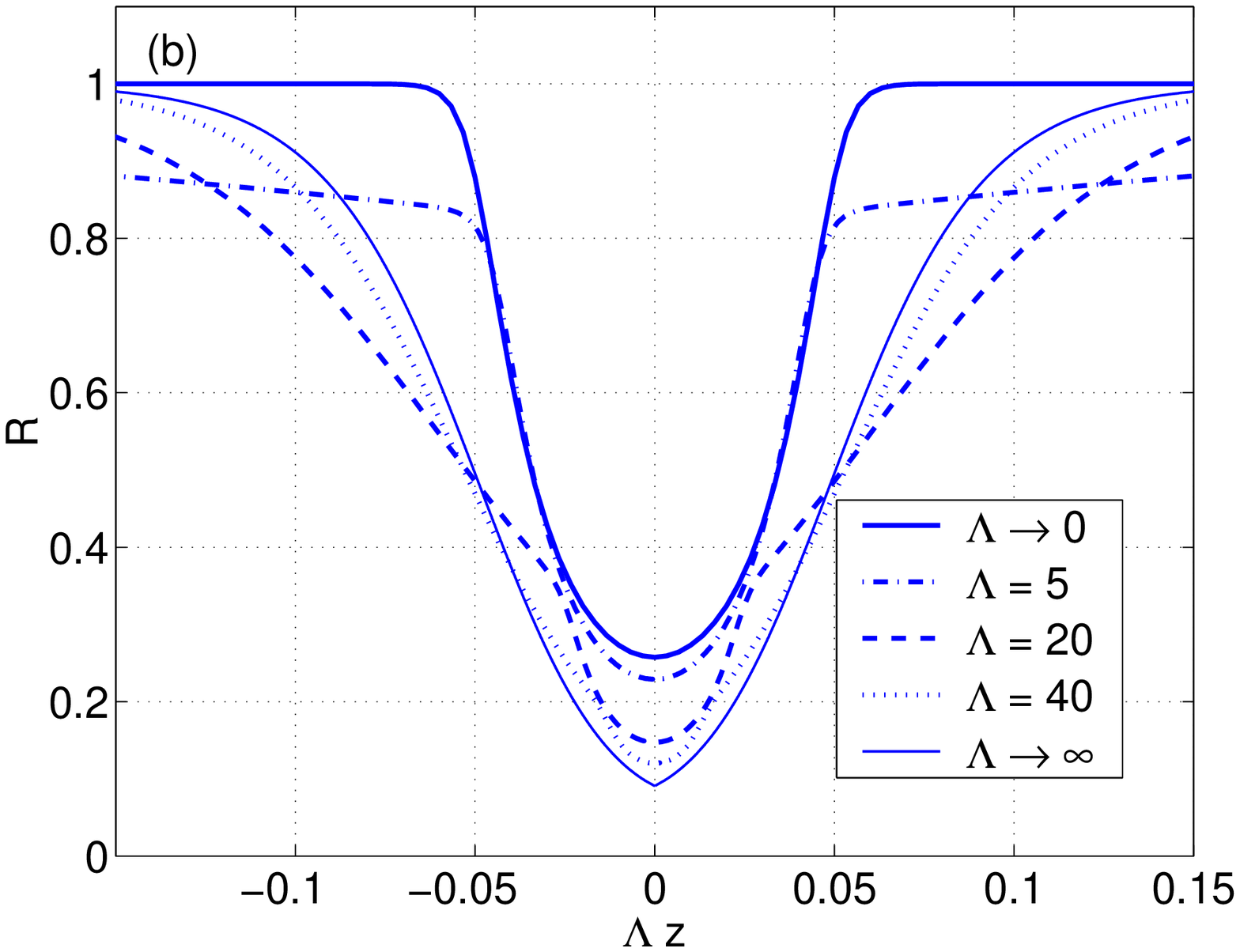}
\includegraphics[width=7.5cm]{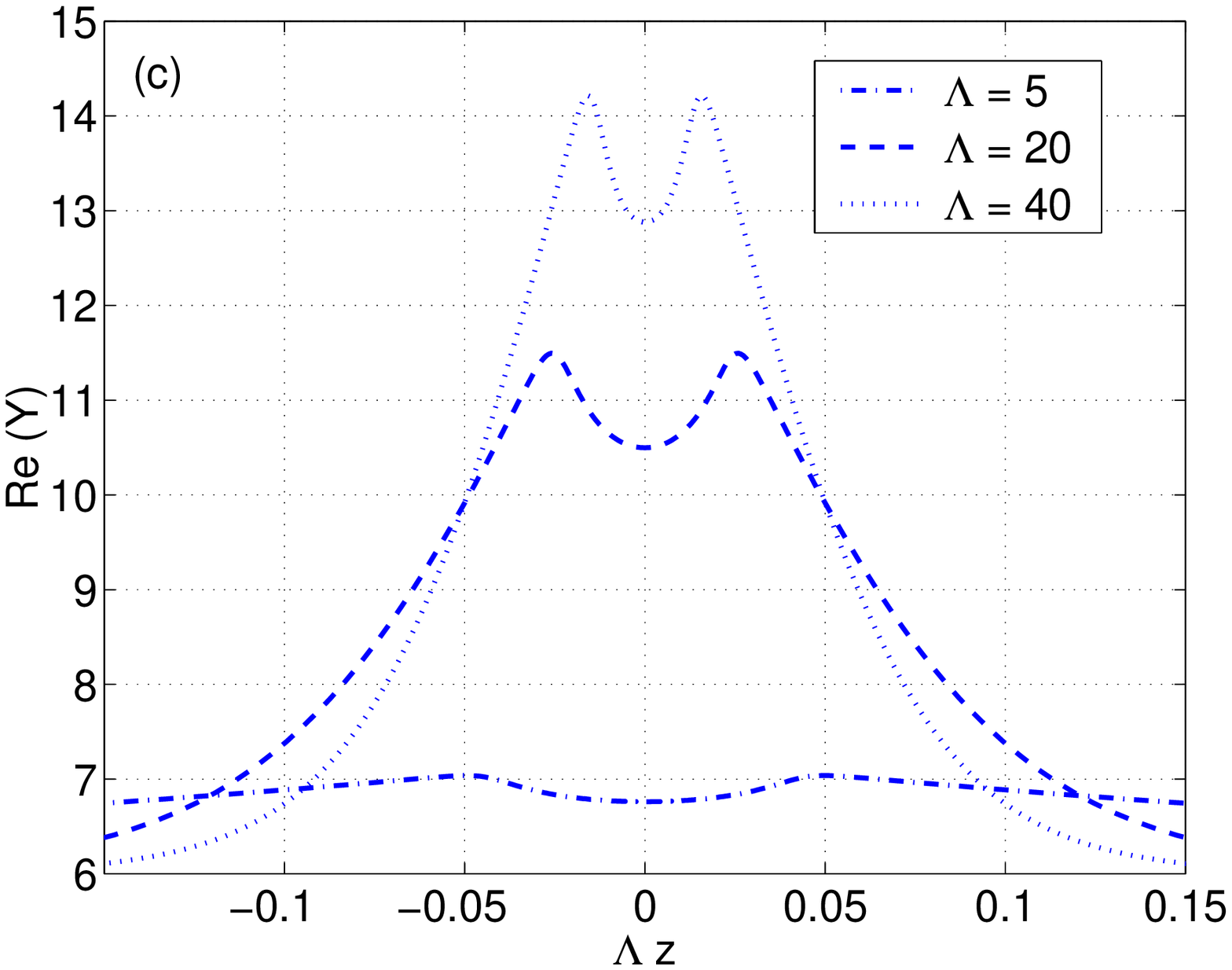} }
\caption{}
{\footnotesize \noindent \textbf
{\bf (a)} Argand diagram of ${\cal W}$ in Model~I, $N=5$, $k=2$ 
for different values of $\Lambda$ (solid: $\Lambda=100$, dot-dashed:
$\Lambda=20$, dashed: $\Lambda=5$); {\bf (b)} profiles of $R$ (absolute 
value of $S$) as a function of $\Lambda z$ for $N=5$, $k=2$ and 
different values of $\Lambda$ in Model~II (dot-dashed:
$\Lambda=5$, dashed: $\Lambda=20$, dotted: $\Lambda=40$). The two
limits are also plotted: VY $\Lambda \rightarrow \infty$ (thin solid)
and LG $\Lambda \rightarrow 0$ (thick solid); {\bf (c)} Profiles of 
${\rm Re} \; [Y]$ as a function of $\Lambda z$ for the same model and 
values of the parameters as before. }
\label{fig4}
\end{figure}

Therefore we have seen that the pattern of domain walls for $k>1$ is
totally different in Models~I and II, although it looks like both
are able to give BPS-saturated domain walls in the VY effective 
limit. In order to clarify this point further let us turn to the
study of the constraint equation, Eq.~(\ref{const}).
     
Let us consider the walls connecting the vacua labelled by integers 
$a$ and $a+k$ (i.e.\ $S(+\infty)
= S(-\infty) e^{ i {2 \pi k}/{N} }$). Using the 
parametrization of Eq.~(\ref{param}), the constraint function for 
Model~I is given by
\begin{eqnarray}
C_{\rm I}(S,Y) &=& R^{1-N}  e^{- {\rm Re} \; [Y] +N-1} \cos \left [(1-N) 
\beta - \frac{\pi k}{N} - {\rm Im} \; [Y]\right] \nonumber \\ 
& + & R \; {\rm Re} \; [Y] \, \cos \left(\beta - \frac{\pi k}{N}\right) 
- R \; {\rm Im} \; [Y]  \, \sin \left(\beta - \frac{\pi k}{N} \right)  -  
N \cos \left( \frac{\pi k}{N} \right) \;\;,
\label{ligBJ1}
\end{eqnarray}
and, for Model~II,
\begin{eqnarray}
C_{\rm II}(S,Y) &=& R^{N+1}  e^{ {\rm Re} \; [Y] -N-1} \cos \left [(N+1) 
\beta - \frac{\pi k}{N} + {\rm Im} \; [Y]\right] \nonumber \\ 
& - & R \; {\rm Re} \; [Y] \, \cos \left(\beta - \frac{\pi k}{N}\right) 
+ R \; {\rm Im} \; [Y]  \, \sin \left(\beta - \frac{\pi k}{N}\right)  +
N \cos \left( \frac{\pi k}{N} \right)\;\;.
\label{ligBJ2}
\end{eqnarray}
The full constraint equations Eqs~(\ref{ligBJ1},\ref{ligBJ2}) are too 
complicated to study in general, but we can still gain some 
information about how the solutions interpolate between domain walls 
in the Landau--Ginzburg models and the VY limit by studying the 
corresponding constraint equations at $z=0$.

We are considering symmetric walls therefore, at the origin, 
$\beta_0 =\pi k/ N$ and ${\rm Im} \; [Y_0]$ = 0, where we use the 
subscript 0 to denote fields evaluated at $z=0$. For Model~I the 
constraint equation takes the form
\begin{equation}
C_{\rm I}(R_0,Y_0) \equiv R_0^{1-N} e^{-Y_0+N-1} (-1)^{k} + R_0 \, Y_0 - N
\cos \left( \frac{\pi k}{N} \right) = 0 \;\;,
\label{ligBJ10}
\end{equation}
and, for Model~II,
\begin{equation} 
C_{\rm II}(R_0,Y_0) \equiv R_0^{N+1} e^{Y_0-N-1} (-1)^{k} - R_0
\, Y_0 + N \cos \left( \frac{\pi k}{N} \right) = 0 \;\;. 
\label{ligBJ20}
\end{equation}
In these equations $Y_0$ is real.
 
We can now plot the relationship between $R_0$ and $Y_0$ for both 
models, as a function of $N$ and $k$. Falling on the constraint curve 
is necessary, but not sufficient, for the existence of a BPS
solution. Figure~5 shows the curve given by Eq.~(\ref{ligBJ10}) for 
$N=5$ and $k=1$ {\bf (a)} and $k=2$ {\bf (b)}, and that given by
by Eq.~(\ref{ligBJ20}) for $N=5$ and $k=1$ {\bf (c)} and $k=2$ 
{\bf (d)}. The horizontal lines represent the value of $Y_0$ for
which the corresponding Landau--Ginzburg ($\Lambda \rightarrow 0$)
limit is reached ($Y_0=N-1$ in the first two plots and $Y_0=N+1$ in
the last two). Note that, in all but Fig.~5b, the constraint equation
intersects the $\Lambda \rightarrow 0$ horizontal line at either one
(Figs~5a,c) or two (Fig.~5d) points.  The intersection at larger 
$R_0$ for Model~II ($N=5$, $k=2$) does not in fact correspond to a 
BPS wall, as is shown in Appendix~\ref{a:c-anal}.

\begin{figure}
\centerline{
\includegraphics[width=7.5cm]{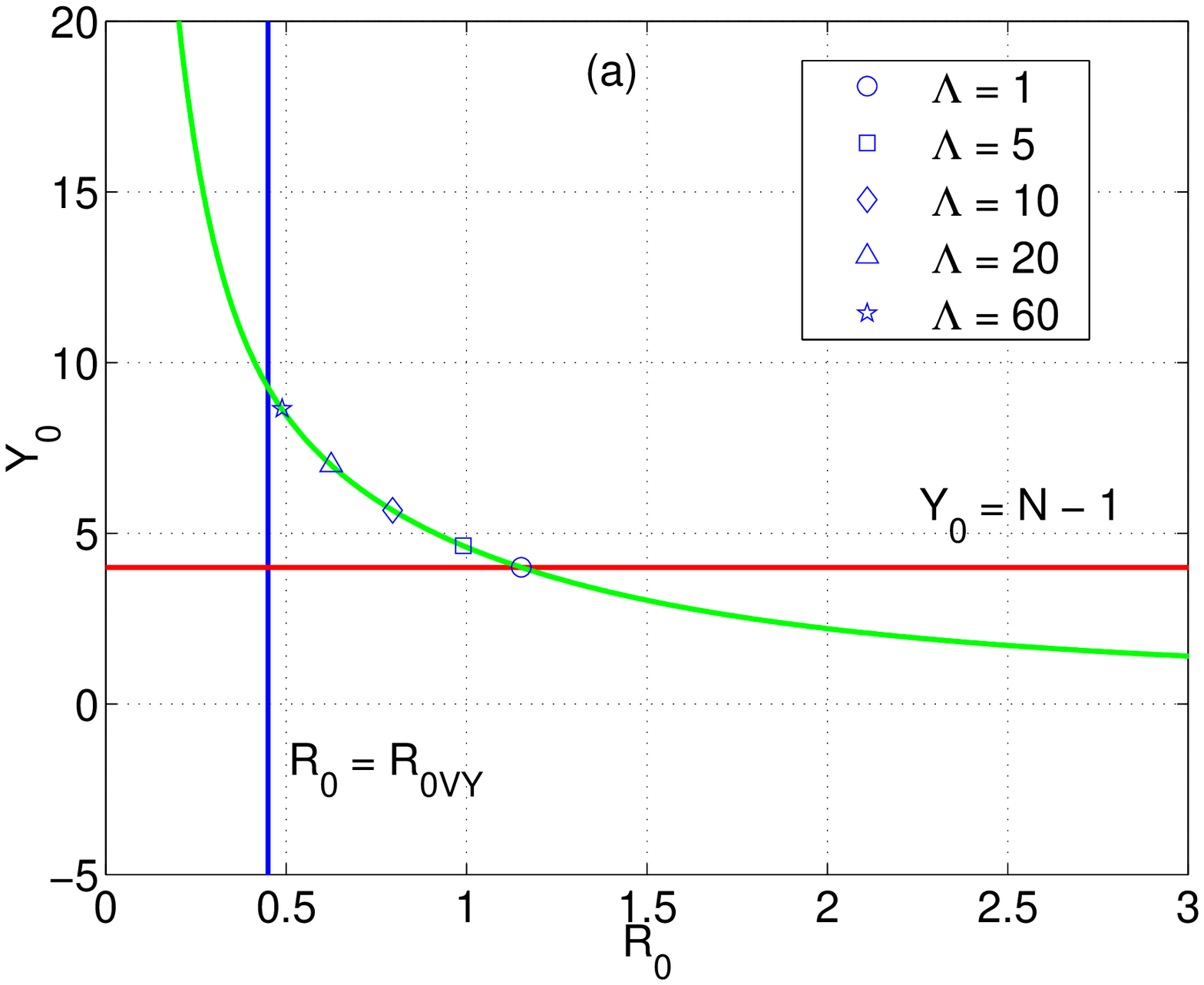}
\includegraphics[width=7.5cm]{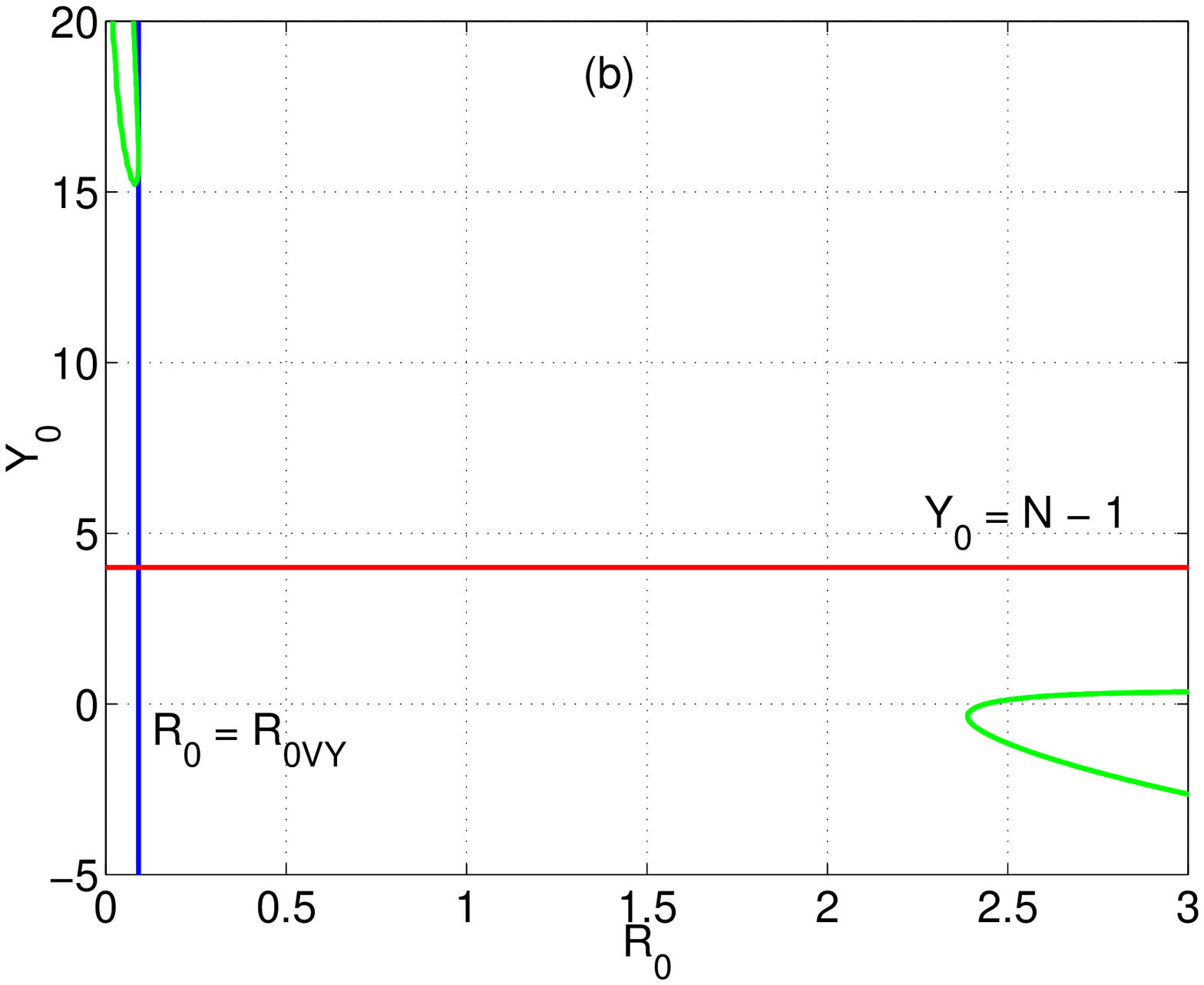} }
\centerline{
\includegraphics[width=7.5cm]{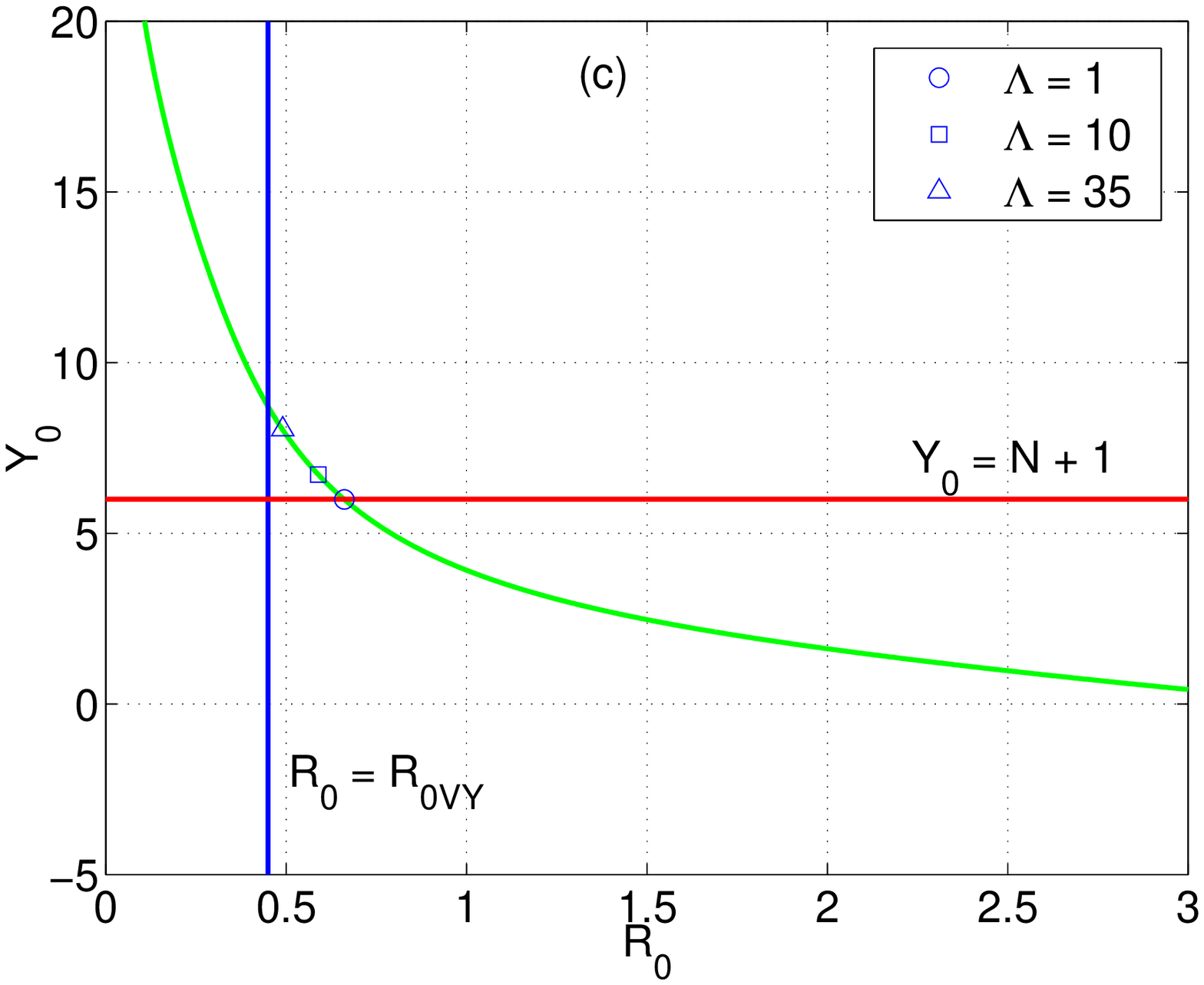}
\includegraphics[width=7.5cm]{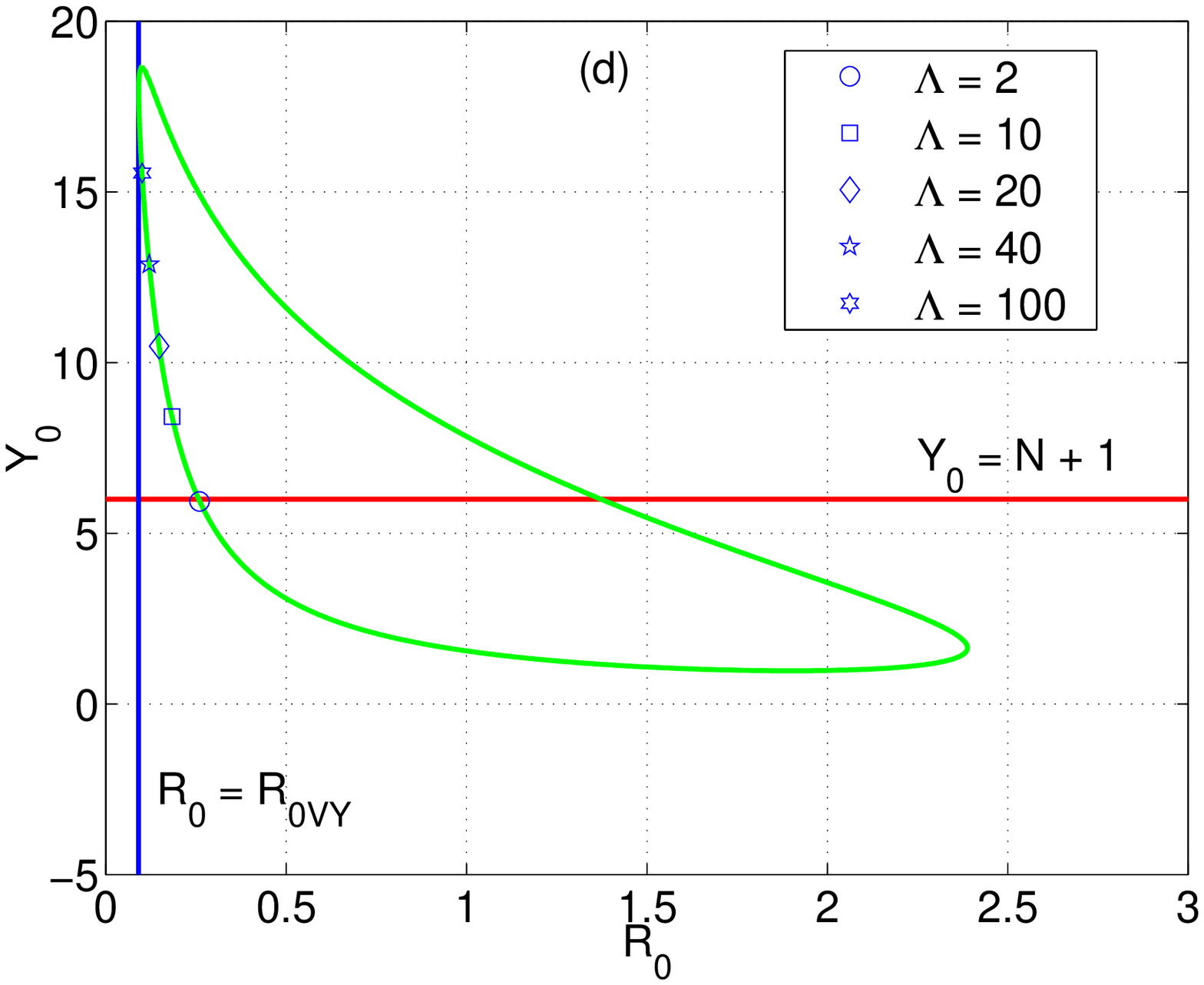} }     
\caption{}
{\footnotesize \noindent \textbf
{\bf (a)} Contour plot of the constraint equation at $z=0$ as a 
function of $R_0$ and $Y_0$ (which is real) for $N=5$ and $k=1$ in Model~I;
{\bf (b)} same for $k=2$; {\bf (c)} same for Model~II, $k=1$;
{\bf (d)} same for Model~II, $k=2$. The horizontal lines represent the
corresponding values for $Y_0$ at the LG ($\La \rightarrow 0$)
limit, whereas the vertical ones represent the values of $R_0$ in the
VY ($\La \rightarrow \infty$) one. The annotated symbols indicate the
BPS-saturated solutions found in each case.}
\label{fig5}
\end{figure}
    
Let us consider now the VY (large $\Lambda$) limit, where the dynamics 
near the minima are governed by the light field, $S$. The $Y$ field 
just follows the evolution of the $S$ field, as dictated by the 
$\W_Y=0$ condition,  $Y=N \mp 1- \ln( {S^N}/{\Lambda^{3N}})$, with the 
minus (plus) sign taken for Model~I (Model~II). However,
this condition must be violated in a region near $z=0$, as it is not 
consistent with ${\rm Im} \; [Y_0] = 0$, $\beta_0 = \pi k/N$. We
denote the width of the region by $\Delta$, such that (with a suitable 
choice of the branch of the logarithm) 
\ben
\ba{ccccc}
Y   &=&   \displaystyle N \mp 1 - \ln \frac{S^N}{\Lambda^{3N}} 
\phantom{ + i 2 \pi k }  &\quad&  z  < - \Delta  \;\;,  \nonumber \\
& & \label{VYlimit} \\     
Y   &=&  \displaystyle  N \mp 1 - \ln \frac{S^N}{\Lambda^{3N}}    
       + i \, 2 \pi k    &\quad&  z  > \Delta \;\;. \nonumber  
\ea
\een
As we will see, in that region $R$ is approximately constant but 
${\rm Im} \; [Y] + N\beta$ changes very quickly to connect the two 
$\W_Y=0$ branches. We would expect the width of the region to be 
governed by the larger of the two masses in the theory, given in 
Eqs~(\ref{e:m2largeL}), in which case it vanishes as $\La^{-3}$ in 
the limit $\Lambda\to\infty$.

In this limit, and using Eqs~(\ref{VYlimit}), the constraint equation 
becomes, 
\begin{equation}
R_0 (1 - \ln R_0) = \cos \left( \frac{\pi k}{N} \right) \;\;.
\label{Rvy}
\end{equation}
This equation has two solutions, one with $R_0 <1$, and the other 
with $R_0>1$.  Our analysis in Appendix~\ref{a:c-anal} 
shows that there is, at most, only one BPS wall for finite $\La$, 
and that it must be the solution with smaller $R_0$. It follows that, 
in the $\La\to \infty$ limit, only one of these solutions
to Eq.~(\ref{Rvy}) corresponds to a domain wall, which is represented 
in Figs~5 by a vertical line, and denoted by $R_{0VY}$. Note also that
the equation is identical to that obtained in 
Ref.~\cite{Decar99} for the effective VY limit, with $k$ replaced by
$N_f$, the number of matter flavours introduced in that case to study
the model in its Higgs phase. There the effective VY limit was
obtained when $m \rightarrow \infty$, with $m$ the mass of the matter
condensates. Again, in that case, only the solutions with $N_f/N<1/2$
(i.e. $R_0<1$) would yield meaningful domain wall profiles, placing a
restriction on the amount of matter one could consider. Here, as we
mentioned before, the fact that it is only possible to define the
effective VY limit for $k<N/2$ does not prevent us from building 
domain walls between {\em any} two neighbours (with the exception of
$N$ even, $k=N/2$).
     
Once we have defined the effective VY limit, let us go back to the 
discussion of our results, for which we focus again on Figs~5. It is 
noticeable how, in all but Fig.~5b, not only do the constraint curves 
intersect the Landau--Ginzburg ($\Lambda \rightarrow 0$) limit, but
also they connect continuously both the Landau--Ginzburg and VY 
($\Lambda \rightarrow \infty$) limits. These are precisely the
curves along which we have found BPS-saturated solutions at finite
$\Lambda$, which are indicated by the annotated symbols, whereas the 
absence of such solutions is obvious in Fig.~5b, where the constraint
curve does not connect the two limits. There is, therefore, a 
straightforward criterion to decide whether a particular model would
yield BPS-saturated solutions for finite $\Lambda$, and that is given
by the fact that the corresponding constraint equation interpolates
continuously between the two existing limits. In the particular
case of Fig.~5d, where there are actually two branches of the
constraint equation that link the VY and Landau--Ginzburg limits, 
it is shown in Appendix~\ref{a:c-anal} that only one of them gives 
BPS domain wall solutions. It is further shown, 
in that same Appendix, that the shape of the constraint curves 
illustrated for $N=5$ and $k=1,2$ in Figs~5 are examples of generic 
classes distinguished by $k < N/2$ and $k$ even or odd, and hence that 
our statements about the number of BPS solutions in the two
models can be extended to all $N$, $k$.

Finally, as mentioned at the beginning of the section, there is the 
possibility, in Model~II, of constructing so-called real domain 
walls, which interpolate between the chirally symmetric vacuum at
($|S|=0$, $Y=0$), and one of the $N$ vacua with $|S| = \La^3$, $Y=N+1$.
Although the chiral vacuum is certainly a supersymmetric minimum of the
theory, in the sense that the scalar potential vanishes there, the 
K\"ahler potential (\ref{e:kahler}) gives a metric which is
singular at that point, so the status of the chirally symmetric 
vacuum is ambiguous. Nevertheless we have verified numerically that 
solutions exist in the model for our choice of K\"ahler metric.

\section{Discussion, physical motivation, relation to previous work}

Once we have shown that both these two Landau--Ginzburg models
admit domain walls, and also have a well-defined effective
VY limit when $\La \rightarrow \infty$, we can proceed to discuss
what the possible physical meaning of the extra $Y$ field should be.
It was already pointed out years ago that the VY approach to SUSY QCD 
was neglecting degrees of freedom which would be as relevant as the
gaugino condensate upon which the VY approach was developed.
For example, spin zero glueballs which, in principle, should not be
heavier than gluino condensates do not contribute to that effective 
action. Therefore it would be justified to attempt a different 
construction which would embed the VY model. This was done
in Ref.~\cite{Farra98} by introducing a real tensor superfield $U$
which contained gluino-gluino, gluino-gluon and gluon-gluon bound
states, and a new $U$-dependent term to extend the VY effective 
action. Therefore this approach preserves the original logarithm and
we do not believe that it could be directly compared to our Models~I
and II.

As was already mentioned in the introduction, the problems
associated with the presence of the logarithm in the original VY model 
were addressed in Ref.~\cite{Burge95} in the context of the linear
multiplet formalism. In that case the authors were focusing on the
study of gaugino condensation in the presence of a field dependent
coupling, which brings into the theory a new scalar field whose 
vacuum expectation value induces the gauge coupling. By studying 
and proving the duality between the linear and chiral formulations
of the problem, they were able to amend the VY chiral potential in
order to eliminate ambiguities coming from the imaginary part of
the gaugino condensate field ($u$ in that case). Again, we have not 
been able to find a direct link between this approach and ours of
extending the theory by introducing a second field $Y$ but,  
nevertheless, we believe that exploring and constructing domain 
walls in this now unambiguous formulation would be a very interesting
exercise.

On the other hand, our $Y$ field certainly has the form of a field 
dependent coupling, as the effective action contains $YS|_F$.  If we 
accept that $Y$ is a coupling field, then it is tempting to interpret 
the extra piece, proportional to $e^{\pm Y}$, as the leading term of 
an instanton-induced contribution.
     
Finally we can comment about the different applications of BPS
domain walls in the context of string theory. According to 't Hooft's  
conjecture~\cite{tHooft}, large-$N$ gauge theories should exhibit a 
phase described by perturbative strings.  It is also 
believed~\cite{Witte97} that, in this limit, BPS-saturated
domain walls are effectively branes on which the QCD string can end.    
Triggered by this, much work has been devoted to studying the 
connection between domain walls and branes in recent years. For
example a construction of BPS domain walls was carried out in 
Ref.~\cite{Dvali99} by exploiting the similarity of the super QCD
Lagrangian (understanding by this the VY model amended to restore the
$Z_{N}$ symmetry~\cite{Gabad99}) with that of the $A_N$ 
Landau--Ginzburg model, in the large-$N$ limit and for walls 
connecting adjacent vacua. In fact, there it was even attempted to  
identify the field content of this super QCD Lagrangian with
those fields appearing in the supersymmetric formulation of the 
membrane action.

There has been also a lot of work on branes and strings at conifold 
singularities. Different dualities are supposed to be related to a 
transition in the geometry. This has been recently suggested by 
Vafa~\cite{Vafa00} in the Chern--Simons/topological strings duality 
framework. He considers Type IIA strings in a non-compact Calabi--Yau
(CY) 3-fold geometry that includes a conifold and adds $N$ D6-branes 
wrapped over the ${\cal S}^3$ in the complex deformed conifold. That 
generates a ${\cal N}=1$ supersymmetric ${\rm SU}(N)$ theory. The 
relevant chiral superfields are $S$, proportional to the one defined 
in Eq.~(\ref{comp}),  and a new chiral field, $Y$, whose lowest
component is given by the volume of the ${\cal S}^3$ (real part) and
by the vev of some 3-form (imaginary part) that will play the role of 
$\theta$ angle. At lowest order, the superpotential is given by
$\W = S Y$ . The new field, Y, triggers the non-perturbative effects. 
Vafa proposes the following corrected superpotential
\begin{equation}
W = \frac{1} \lambda_s S Y + i N^2  e^{-Y/N} \;\;,
\end{equation}
where $\lambda_s$ is the string coupling . If we integrate out the $Y$
field, by using $\W_Y = 0$, we end up with the VY superpotential for the
condensate. Notice that this superpotential shares two properties with
our models: there is a linear, $ S Y$, term and an exponential term
involving the new field, $Y$. The combination of these two facts 
gives the logarithmic dependence in the effective Lagrangian 
describing the gaugino bilinear. Therefore we believe that our models
could shed some light on the role played by the fields that
couple to the gaugino condensate. These fields are very important
if we want to understand the core of the domain wall.

\section{The large-$N$ limit } 

We have seen in the previous examples how to calculate BPS domain
walls for the VY Lagrangian in the cases where they exist in the
extended theory. Let us assume that the complete theory supports
these BPS domain walls. Now we can study how they behave
for large values of $N$ (see also Refs~\cite{Dvali99, Gaba99, Kaku99}).
This is particularly important because, in that 
limit, these domain walls are thought to be the field theory 
realization of branes. 

We want to solve the BPS equation 
\begin{equation}
{\cal K}_{S\bar{S}} \partial_z \bar{S}   =  e^{i\gamma} \ln 
\frac{S^N}{\Lambda^{3N}} \;\;,
\label{VY}
\end{equation}
for the $(N,k)$ case (i.e.  $\gamma = \pi /2  - \pi k/N $), and 
large $N$ values. In order to do that, it is useful to define the 
parameter $ \epsilon = \frac{ k \pi}{N}$. Then the constraint for 
$R_0$, Eq.~(\ref{Rvy}), becomes
\ben
\cos \epsilon - R_0 \, (1-\ln R_0 ) = 0 \;\;.
\label{R0}
\een
Notice that the allowed values of $k$ will depend on the details
of the theory that generalizes the VY Lagrangian. For example,
in Model~I only $k=1$ will be allowed\footnote{A similar result was 
found in Ref.~\cite{Kaku99}, where the large-$N$ limit is mapped into
a Landau--Ginzburg model.}, whereas in Model~II we can consider any 
value of $k$. In any case, assuming $ k \ll N$, we can expand $R_0$ in 
terms of $\epsilon $ to obtain its large-$N$ limit 
\ben
R_0 =  1 \pm \epsilon + \frac  {\epsilon^2}{6} + {\cal O} 
(\epsilon^3) \;\;.
\een
Then $|S|$ deviates from its value at the minima by ${\cal O} 
(\epsilon)$. Since the phase of the condensate is also ${\cal
O}(\epsilon)$, we will adopt the following parametrization
\ben
S(z) = \Lambda^3 ( 1 - \epsilon \rho(z) ) \, e^{i \epsilon \psi(z)} \;\;,
\een
with  $\rho(z),  \psi(z)$ real functions of ${\cal O}(1)$. Substituting 
the previous expression into the BPS equation, Eq.~(\ref{VY}), we get 
\begin{eqnarray}
\rho(z)  & = &  e^{9  z/\Gamma} + (e^{9  z/\Gamma} - \frac{7}{6} 
e^{18  z/\Gamma} ) \epsilon +   {\cal O} (\epsilon^2) \;\;, \nonumber
\\
& & \label{sysN} \\     
\psi(z)  & = &  e^{9  z/\Gamma } +  {\cal O} (\epsilon^2) \;\;, \nonumber
\end{eqnarray}
for the $z<0$ branch (and the `symmetric' one in the $z>0$ branch).
In these equations $\Gamma =  \frac{1}{\Lambda N}$ fixes the domain wall
width. In order to determine the $N$ dependence of the relevant 
quantities (i.e. width and tension of the wall) in the large-$N$ limit,
it is standard to rescale the fields so that the Lagrangian scales like
$N$. For example, Eq.~(\ref{cond}) will appear as
\begin{equation}
\langle \lambda \lambda \rangle_k \equiv \langle {\rm Tr} \lambda^a
\lambda_a \rangle_k = N {\tilde \Lambda}^3 e^{i \frac{2\pi}{N} k} \;\;.
\label{cond2}
\end{equation}
If we calculate now the domain wall tension when interpolating from
vacuum $a$ to vacuum $a+k$ we get
\begin{equation}
\sigma =  |W_{a+k}-W_a| =  N^2 {\tilde \Lambda}^3 2 \sin \epsilon  
\;\;,
\label{2field}
\end{equation}
i.e. for $k = {\cal O}(1)$ the tension is expected to scale
as $N$. Notice that there are two different ways to calculate
the domain wall energy: using the two-field theory or the effective VY
one. The validity of  Eq.~(\ref{2field}) relies on the 
differentiability of the domain wall profile, so it can be only used 
in the two field theory. In the VY model, one should split the energy 
into the contributions coming from the $z<0$ and $z>0$ branches. 
In that case we get
\begin{equation}
\sigma_{VY} =   N^2 {\tilde \Lambda}^3 2  | \sin \epsilon -  \epsilon R_0|
\;\;.
\label{1field}
\end{equation}
This result coincides with the one obtained by using squark matter 
fields (see Ref.~\cite{Smilg01}) as the extra fields providing regular 
domain walls. This is due to the fact that, in both cases, the same 
prescription is used to calculate the imaginary part of the logarithm.
We have seen how this prescription naturally emerges when integrating 
out the $Y$ field (see Eq.~(\ref{VYlimit})), as it was also shown in 
Ref.~\cite{Decar99} for the squark matter fields. Notice, however, 
that for $k>1$ the squark fields, as described by the TVY Lagrangian, 
do not provide regular BPS domain walls for high enough values of 
their masses. This means that, in these cases, the extra fields that 
enter the logarithmic term in the effective potential can not be
ignored. They play a relevant role and are excited in the subcore of 
the domain wall. 
 
On the other hand, the difference between $\sigma$ and $\sigma_{VY}$ 
gives us information about the energy that is stored in the $Y$ subcore, 
i.e. the spatial region in the large $\Lambda$ limit where the phase 
of $Y$ changes quickly. Since  $\sigma  \sim {\tilde \Lambda}^3 N $ 
and $\sigma_{VY} \sim {\tilde \Lambda}^3 $, we conclude (as first shown 
in Ref.~\cite{Smilg01}) that, in the large-$N$ limit, most of the 
energy concentrates in this subcore.

\section{Conclusions}
     
We have studied the existence of BPS-saturated domain walls in two
extensions of the Veneziano--Yankielowicz effective Lagrangian 
formulation in order to study SUSY Yang--Mills theories. These are motivated 
by the existence of several problems associated with the presence of
a logarithm of the gaugino condensate field in the effective potential 
of the VY formulation, which causes ambiguities when studying
dynamical questions such as the formation of domain walls. Our 
extensions, which we denote as of the Landau--Ginzburg type, avoid such 
problems by introducing an extra chiral superfield, $Y$, and a
logarithm-free interaction which, in the limit of heavy $Y$,
result in the effective VY model. Within this framework we have 
constructed domain walls which turn out to be BPS-saturated in one of the
models but not in the other one. In any case, both seem to give 
BPS-saturated walls in the effective VY limit.
     
We have also studied the limit of light $Y$ field, in which the
models proposed approach Landau--Ginzburg models with which one can
construct well-defined BPS-saturated walls. It turns out that  
it is possible to define a criterion to decide whether our
particular extensions of VY admit BPS-saturated walls at any point in
between the two (VY and LG) limits: this is given by whether the BPS 
constraint, associated with the corresponding BPS equations,
is able to interpolate continuously between both limits. Our results
can be proven to hold for any values of $N$ (number of colours) and 
$k$ (neighbours between which we interpolate).
     
Finally we comment on the possible physical significance of the
extra field $Y$, and the connection of our approach with other
proposed models existing in the literature. Although there is no
direct evidence so far, the similarity of this model with some
domain wall constructions done in the context of string theory is quite
remarkable, and it deserves further investigation.

\section*{Acknowledgements}
    
BdC would like to thank Jean-Pierre Derendinger, Nick Dorey, Tim
Hollowood and Prem Kumar for very interesting suggestions and
discussions. JMM thanks C\'esar G\'omez for making very useful 
suggestions. The work of JMM is supported by CICYT, Spain, under 
contract AEN98-0816, and by EU under TMR contract ERBFMRX-CT96-0045,
and that of BdC, MBH and NMcN is supported by PPARC. Both BdC and
JMM would like to thank the CERN Theory Division for hospitality 
during intermediate stages of this work.

\appendix
   
\section{The Newton method for finding solutions}
\label{a:numerics}

The numerical solutions to the field equations were found with 
Newton's method, also called the Newton-Raphson method (NR). This was 
implemented by a Fortran 90 program.

We wish to find extremal energy configurations for the discretized energy 
functional of the domain wall, $E(f_A)$, where $f_A$ 
denote the degrees of freedom: here, the values of the complex fields 
and their conjugates at each point on the lattice.

The Newton-Raphson algorithm consists of iterating the update
\bea df_A = - \left( \frac{{d^2}E}{{df_A}{df_B}}(f) \right) ^{-1} 
\frac{dE}{df_B}(f) \;\;. \label{matr} \eea
In order to solve the matrix equation (\ref{matr}) our program calls  
linear algebra routines {\sc zgbtrf} and {\sc zgbtrs}
from the Silicon Graphics implementation of the BLAS library. 

It is necessary to use a rescaled lattice because of the presence of widely 
differing mass scales when $\Lambda$ differs from $O(1)$. The rescaling used 
takes the form
\bea s = \half ( \tanh({m_1}{x}) + \tanh({m_2}{x}) ) \;\;, \eea
where $x$ is the physical position, $s$ is the rescaled position and
$(m_1,m_2)$ are parameters chosen to give the best spread of lattice 
points. There should be enough points in the central region to 
accurately follow the rapidly changing fields, while the total number 
of points should be minimized so that the program takes less time to run.

Once the rescaling is chosen the lattice points are placed equally 
spread in $s$ between $s = -1$ and $s = 1$. This means that the end 
points of the lattice are at spatial infinity.

The number of lattice points used was typically between 100 and 300, and the 
number of iterations needed to reach convergence was between 20 and 1000.

Spatial derivatives are calculated three ways, with forward, backward and 
symmetric derivatives. 
The energy function $E(f_i)$ has contributions from all three forms of 
the derivative, added with equal weighting. The advantage of this 
approach is that symmetry between left and right is maintained, while 
odd and even numbered points are kept in close contact, which would 
not be the case were only symmetric derivatives used.

The values of the fields on the boundaries are set equal to their 
vacuum values, and are kept fixed. The walls are symmetrical between 
left and right, because of the $Z_N$ symmetry in the Lagrangian. This 
links the field values in one half of the wall with the 
conjugates of the values in the other half, i.e.
\ben
S(s) = \bar{S}(-s) e^{i 2 \pi k / N},\quad
Y(s) = \bar{Y}(-s) \;\;.
\een
However, rather than model half of the wall, it was decided to model 
both sides, because of the difficulty of establishing simple boundary 
conditions at $z=0$, consistent with the strategy of programming with 
complex, rather than real, fields. 

The convergence criterion was framed in terms of $c = \sum_A \left| dE/df_A 
\right|$. The algorithm is judged to have converged when $c$ reaches $\lap 
10^{-10}$.

When far from a solution the size of the update may have to be reduced. It is 
guaranteed that, for a sufficiently small movement, following $-df_A$ will 
approach a solution. However, a full NR update may overshoot, and lead to 
divergent field values, rather than converging. To deal with this, in the 
initial stages of a run the size of the update is reduced by a factor
of 10, or even 100. Systematic methods of doing this, such as 
backtracking along $df_A$, were judged to be unnecessary.

To check the BPS saturation of a solution two criteria are
used. Firstly, the total energy of the wall is compared to the BPS 
energy, which is itself easily calculated. Because of discretisation 
error, the numerically determined energy is usually below the true BPS value.

The second method of determining whether solutions are BPS is to 
check that the wall follows a straight line in $W$-space.
The straightness of the line is quantified by the ratio of the 
maximum deviation from a straight line linking the ends, divided by 
the length of the line. If this quantity can be shown to decrease 
towards zero as the number of lattice 
points is increased, the configuration is taken to be BPS.

\section{Detailed analysis of the BPS constraint equations}
\label{a:c-anal}

In this Appendix we supply details of the analysis behind 
Section~\ref{s:dwconstruct}, where we described the values of $N$ 
and $k$ admitting BPS domain walls. 

In Appendix~\ref{a:LGlimit} we study
the limit $\La \to 0$, where our models tend 
to the Landau--Ginzburg models LG1 and LG2 described in
Section~\ref{s:models}.
In this limit we are able to prove exact results giving the values of 
$(N,k)$ for which BPS walls exist and, conversely, those for which BPS 
walls do not exist.  Our argument uses continuity in $\Lambda$ to 
make the statement that BPS walls cannot exist in a particular model at
$\Lambda > 0$ if its $z=0$ 
constraint curve, Eqs.~(\ref{ligBJ10}, \ref{ligBJ20}), reaches
the LG limit at a point where we know there is no BPS domain wall.  

To this end, we must analyse the $z=0$ curves, to enumerate all possible 
BPS domain wall solutions, and to check that they have an LG limit.  This 
is done in Appendix~\ref{a:const0}.

\subsection{Analysis in the Landau--Ginzburg limit}
\label{a:LGlimit}     

In the Landau--Ginzburg limit, $\Lambda \to 0$, the extra field $Y$ stays 
at its vacuum value $Y_*=N \mp 1$.  We require that solutions of the 
constraint equations Eqs~(\ref{ligBJ1},\ref{ligBJ2}) exist for all
$\be$ in the range $0 \le \be \le 2\pi k/N$, and asymptote to $R = 1$ 
at $\be = 0$ and $\be = 2\pi k/N$.  Any solution to the constraint 
equation at $z=0$ which does not connect the vacua is a `fake' 
solution and cannot represent a BPS domain wall. By continuity in
$\La$, solutions to the full constraint equation for our interpolating 
models which tend to these fake solutions as $\La \to 0$ are also 
inadmissable as BPS walls.

Firstly, consider the Model~I constraint equation Eq.~(\ref{ligBJ1}), 
and define a new angle $\de = \be - \pi k/N$, so that the constraint 
function becomes
\ben
C_{LG1}(R,\de) = R^{1-N} \cos[(N-1)\de](-1)^k + (N-1) R \cos\de - 
N\cos \left(\frac{\pi k}{N} \right) \;\;.
\een
Where $\cos[(N-1)\de](-1)^k > 0$, the constraint function has a single 
minimum as a function of 
$R$ at $R=R_{\rm min}$. One can show that as one varies $\de$, the
maximum value of 
this minimum occurs at $\de_{{\rm max},m} =\pi m/N$, with $-k \le m \le 
k$, and $m$ even (odd) when $k$ is even (odd).  At these saddle points 
\ben
C_{LG1}(R_{\rm min},\de_{{\rm max},m}) = N \left(\cos \left( 
\frac{\pi m}{N} \right)  - \cos \left( \frac{\pi k}{N} \right) \right) \;\;.
\een
When this is greater than zero there can be no solutions to 
$C_{LG1}(R,\de) = 0$. For $k>1$, there is always a value of $m$ for which 
$C_{LG1}(R_{\rm min},\de_{{\rm max},m}) > 0$, and
so the constraint curve cannot move continuously between $\de = 0$ and 
$\de = \pi k/N$.   

Hence we conclude there are no BPS domain walls in the
Landau--Ginzburg limit ($\La \to 0$) of Model~I for any $k>1$, and
only one for $k=1$

For the Landau--Ginzburg limit of Model~II, we can rewrite the constraint
function as
\ben
C_{LG2}(R,\de) = R^{N+1} \cos[(N+1)\de](-1)^k - (N+1) R \cos\de + 
N\cos \left(\frac{\pi k}{N} \right) \;\;.
\een
Where $\cos[(N+1)\de](-1)^k > 0 $ the constraint function has a single 
minimum at $R=R_{\rm min}$. One can again show that the maximum value
of this minimum occurs at $R_{\rm min} = 1$, $\de_{{\rm max},m} = \pi
m/N$, with $-k \le m \le k$, and $m$ even (odd) when $k$ is even
(odd). At these saddle points  
\ben
C_{LG2}(R_{\rm min},\de_{\rm max}) = N \left(\cos \left( 
\frac{\pi k}{N} \right)  - \cos \left( \frac{\pi m}{N} \right) \right) \;\;.
\een
In this case, the value of the constraint function at the saddles is 
always less than zero (unless $m=k$), and therefore there are two 
solutions to the equation $C_{LG2} = 0$ when $\cos(\pi k/N) \ge 0$,
and only one if $\cos(\pi k/N) < 0$. Where $\cos[(N+1)\de](-1)^k < 0$, 
the constraint function becomes a monotonically decreasing function of 
$R$. For these values of $\de$ there can only be one solution to the 
constraint equation for $\cos(\pi k/N) \ge 0$, and none for 
$\cos(\pi k/N) < 0$. Hence only one of the solutions to the constraint 
equation at $\de=0$ (which one can straightforwardly see is the
one with smaller $R$) can
connect two vacua for $\cos(\pi k/N) \ge 0$, and none for $\cos(\pi k/N) 
< 0$. 

In fact, we can also eliminate the solution at $R=0$ for 
$\cos(\pi k/N)=0$, as it cannot interpolate between vacua with $R=1$.  

Hence we conclude there are no BPS domain walls in the 
Landau--Ginzburg limit ($\La \to 0$) of Model~II 
for any $k \ge N/2$, and only one for $k< N/2$

\subsection{Analysis of constraint curves at $z = 0$}
\label{a:const0}

In this Appendix we show that the constraint curves at $z=0$ fall into 
a small number of classes: four if $N$ is odd, and six if $N$ is even.
These classes are distinguished by having the same number of
intercepts with the lines representing the LG and VY limits, and by 
having the same number of points at which they are either horizontal 
or vertical. This classification helps us to be certain that we have 
found all possible types of BPS wall solution in Section~\ref{s:dwconstruct}.

Firstly we note that, as $k$ is an integer, $\cos(k\pi)$ can only 
have the values $1$ or $-1$ and, if $N$ is odd, $\cos(k\pi/N)$
can be either positive or negative. There are therefore 
four distinct cases for which we can expect the constraint curve to be 
qualitatively different. If $N$ is even, $\cos(k\pi/N)$ can be zero, and 
there are therefore six classes.

We have already eliminated $\cos(\pi k/N)\le 0$ from consideration in the 
Landau--Ginzburg limits of both models, and thus we need concentrate
only on the cases $\cos \pi k = \pm 1$, with  $\cos(\pi k/N) > 0$
For illustration purposes it is therefore 
sufficient to consider, as we have done, 
$N=5$ with $k=1,2$ only. 
     
We look for intercepts with four particular curves, namely
\bea 
R_0 &=& 1 \;, \label{line1} \\
Y_0 &=& 0 \;, \label{line2} \\
Y_0 &=& N \mp 1 \;,\label{line3} \\
Y_0 &=& N \mp 1 - N\ln R_0 \;, \label{line4} 
\eea
with the negative sign taken for Model~I and the positive sign for 
Model~II. Note that the third of these is the LG limit, and the fourth 
is the VY limit in the case of even $k$. Along these curves the 
constraint equations are sufficiently simple that the number of 
solutions can easily be deduced. 

The other information we use is the slope of the constraint curves
\ben \frac{\partial Y_0}{\partial R_0} = \left\{
\ba{cr}\displaystyle
-\frac{(N - 1) R_0^{-N} e^{-Y_0+N-1}(-1)^k  - Y_0} 
{R_0 (R_0^{-N} e^{-Y_0+N-1} (-1)^k - 1)} & \textrm{Model I} \\[15pt]
\displaystyle -\frac{(N + 1) R_0^{N} e^{Y_0-N-1}(-1)^k  - Y_0} 
{R_0 (R_0^{N} e^{Y_0-N-1} (-1)^k - 1)} & \textrm{Model II}
\ea
\right.
\een
By noting the sign of $\pa Y_0/\pa R_0$ we can trace the shape of 
the curve. Note that for even $k$ the constraint curve is vertical at 
$Y_0 = N \mp 1 - N\ln R_0$, which is precisely at the VY limit.  This 
is illustrated in Figs~5b and 5d, where it can be verified that the 
constraint curves are vertical at $R_0 = R_{0VY}$.  

Now we shall consider the two cases ($k$ even or odd) in detail, in Model I.

When $k$ is even (see Fig~5b), Eqs.~(\ref{line1},\ref{line3}) can not 
be satisfied, so the constraint curve never crosses the lines 
$R_0 = 1$, $Y_0 = N \mp 1$. However, it does cross Eq.~(\ref{line2}) 
once at $R_0 > 0$ and Eq.~(\ref{line4}) twice, once on either side of 
$R_0 = 1$.

Let us trace the curve from its $R_0 <1$ intercept with 
Eq.~(\ref{line4}). For the part of the curve with $Y_0 > N-1 - N\ln
R_0$, the slope is negative, and so the curve must move left and
up. The part with $Y_0 < N-1 - N\ln R_0$ has positive slope but, 
because it cannot cross $Y_0 = N-1$, it must reach a turning point,
and also move left and up.  This accounts for the upper branch of the 
curve in Fig.~5b.

Starting now from the $R_0>1$ intercept with Eq.~(\ref{line4}), the 
lower part of the curve will follow $R_0\rightarrow\infty, 
Y_0\rightarrow-\infty$. The upper part will cross Eq.~(\ref{line2}) 
and then $\pa Y_0/\pa R_0 = 0$ and so turn down again. It cannot cross 
Eq.~(\ref{line2}) again, and so $Y_0\rightarrow 0$ from above, as 
$R_0\rightarrow\infty$.

When $k$ is odd, (Fig~5a), we can by inspection of Eq.~(\ref{ligBJ20})
see that $C_{\rm I}(R_0,Y_0) = 0$ only when $Y_0 > 0$. 
Furthermore Eqs.~(\ref{line1},\ref{line3}) can each be satisfied 
once, while Eq.~(\ref{line4}) is never true: hence the curve is never 
vertical, and in fact the slope is always negative. This means that 
$Y_0\rightarrow 0$ from above as $R_0\rightarrow\infty$, while as 
$R_0\rightarrow 0$, $Y_0\rightarrow\infty$.

One can go through a similar analysis to satisfy oneself that the
curves for Model~II are similar to Figs.~5c, 5d. The principle 
differences between the models occur for even $k$ and can be seen by 
comparing Fig.~5b to Fig.~5d: Model~I has no intercepts with $R_0=1$, 
while Model~II has two. There are also no intercepts with $Y_0 = 
N-1$ in Model~I, but one can easily check that there are two with the 
analogous line $Y_0=N+1$ in Model~II. The intercept with $R_0 >1$ does 
not however correspond to a domain wall in the Landau--Ginzburg 
($\La \to 0$) limit, as was shown in Appendix~\ref{a:LGlimit}.

\end{document}